# Logical contradictions of Landau damping

V. N. Soshnikov[1]

Plasma Physics Dept.,
All-Russian Institute of Scientific and Technical Information
of the Russian Academy of Science
(VINITI, Usievitcha 20, 125315 Moscow, Russia)

## Abstract

Landau damping/growing at boundary condition of excitation $\sim \cos\omega t$ of a harmonic wave in collisionless ion-electron-neutrals plasma contradicts to the law of energy conservation of a wave damping/growing in space. There is also no criterion of a choice either damping or growing solution in difference from always non-damping in the direction of their propagation Vlasov waves. Variety of other incongruities as consequence of Landau damping is specified also. Absence of explicit positivity and finiteness of wave solutions for electron distribution function near singularity point leads to need of imposing additional cutting off constraints with resulting positivity and finiteness of the electron distribution function at the singularity point and finiteness of the complex dispersion integral. Landau damping as a real physical phenomenon of collisionless damping does not exist. A relation is established for the real dispersion equation with real waves (see Appendices 2, 4) between the averaged over period wave damping decrement and the collisional term of kinetic equation. Collisionless Vlasov -- Landau damping is explained finally by the usual wrong use of nonlinearly complex wave functions leading to complex dispersion equation. All used solution of the complex dispersion equation for the simultaneously existing collisionless both exponentially damping and growing complex waves is entirely, quantitatively and in its logical sense, different from the solution of initially real dispersion equation for real either damping or growing waves and should be discarded (see Appendices 2, 4, 5, 6). Collisionless damping is caused by unreasonable use of wave functions with complex frequency or complex wave number leading to complex dispersion relation with nonphysical binomial complex roots out of connection with energy conservation law. But all this transfers the finding such virtual roots of the complex dispersion equation with such waves to the field of only abstract mathematical interest.

The origin of a tremendous amount of works on the wave propagation and evolution of plasma perturbations is the past century work [1], where the problem was formulated as a solution of two coupled equations that is collisionless kinetic and Maxwell equations for charged plasma particles.

The typical feature of these equations of collisionless ion-electron-neutrals plasma with the complex wave functions $\sim E_0 \cdot \exp[i(\omega t - kx)]$ is appearance in dispersion equation the indefinitely divergent integrals (IDI) with logarithmic divergence. Here $\omega$ is wave frequency, $k$ is wave number, $E(x,t) \sim iE_0 \exp[i(\omega t - kx)]$ is perturbed longitudinal electrical field at exiting external field $E(x,t)$ in the point $x=0$. At real $\omega$ and $k$ these wave functions lead to the real dispersion equation [3]. Later this solution was generalized also for transversal plasma waves. As it is well known, in the work [1] these integrals along electron velocity $v_x$ were implied in principal value sense, but eight years later ones have taken Landau proposition [2] to imply these integrals as contour integrals with going out into the complex plane of the real value integrand variable $v_x$ and the contour bypassing around the poles of $v_x$ at real axis $v_x$ below or above, which results in complex dispersion equation with complex $k = k_1 - ik_2$ and the so-called Landau exponential damping (or growing) effect, which now has not to be subjected to any doubt and critical discussion.

It can appear however that emerging of IDI in equation solutions is directly a result of information lack (defects) in the original equations. It means that the sense of IDI should follow by single way some additional physical conditions (missing in incomplete collisionless kinetic equation) which are specific for each of the problems. If one follows this principle, strictly speaking, understanding the IDI *ad hoc* either in the sense of principal value or contour integral should be equitable in the case of complex dispersion equation $\omega = f(k)$ obtained using complex wave functions with, for example, real electron wave frequency $\omega$, as further in this work, and complex wave number $k = k_1 - ik_2$, when complex dispersion equation with the imaginary value of the residue in a singularity point pole (see [3]) leads to damping/growing of the electron wave in a collisionless plasma (Vlasov-Landau damping/growing, cf. below *Appendix 2*, what allows to combine both authors in the name of this effect).

---

[1] E-mail: vikt3363@yandex.ru

The method of solving the wave dispersion equation is Laplace transform particularly used in [2] for obtaining the approximate solution of complex dispersion equation. But in fact, the problem is reduced in existing academic and educational literature to a particular way of solving complex dispersion equation obtained by extrapolation of the real dispersion equation by substituting the real $k$ for the complex $k \to k_1 - ik_2$. Crucial issue when using complex wave functions with complex dispersion equation is its solution $k_1, k_2 = f(\omega)$ as a complex root unrelated to the energy conservation law with its violation at $k_2 \neq 0$ (with non-damping Vlasov waves at $k_2 = 0$). In *Appendix 2* and the following Sections it will be detailed argued fallacy of damping/growing of plasma waves in collisionless plasma in the sense of lack energy-exchange right hand side terms of electron kinetic equation.

There are also analyzed below some simplest modifications of the plane boundary problem with the given perturbation boundary sinusoidal electrical field $E(x=0, t)$ for half-infinite and infinite homogeneous column of electron plasma as a simple illustration of questions of principle [4]. It is supposed however for specify setting the boundary problem that the electron velocity distribution function $f_0(v, v_x)$ is Maxwellian $f_0(v)$ at $x=0$, that is either at both sides $x = \pm 0$ (thus the plasma is filling the whole unclosed volume $x \to \pm \infty$), or otherwise the plane $x=0$ is ideally electron reflecting wall. The symmetry or anti-symmetry of the waves relative to the region of their excitation $x \sim 0$ is determined by the way of their excitation.

In this case the additional physical conditions are both absence of backward traveling wave of the half-infinite wave solution (owing to the absence of any sources of reflection or exciting backward waves in remote parts of homogeneous plasma) and absence of the so-called kinematic waves arising at independent of the boundary field excitations of electron distribution function [5]. In plasma of charged particles such kinematic waves are impossible without by them produced changes of electric field including the already given periodical boundary field, otherwise speaking the given boundary field must hardly determine the missing boundary values of perturbation of electron distribution function $f_1 \equiv f_1(v, v_x)\cos(\omega t - kx)$ at $x=0$ and the full approximate electron distribution function $f = f_0(v) + f_1(v_x)\cos(\omega t - kx)$ (where $f_1(v_x) \equiv f_1(v, v_x)$).

At propagation of transverse waves there are now two coupled non-damping (in according to Vlasov longitudinal waves) modes, the forward fast electromagnetic ("usual") wave and the low speed electron wave mode. If the waves are excited by a single pulse of the boundary electric field, then an effect arises of very weak back response ("echo"), thus its explanation in this problem should be not related with Landau damping, possibly with phase interactions of resonant modulated electron beams (hypothetical Van Kampen modes) and nonlinear wave interactions (with nonlinear terms in the original equations) [6].

The single-valued solutions which have been obtained with IDI in principal value sense satisfy both the original equations and the abovementioned additional physical conditions. It is shown that in the case of the integration contour in complex plane $v_x$ these additional physical boundary conditions can not be fully satisfied [4].

As a verifying of Landau damping, ones indicate few experiments with the frequency dependence of wave decrement resembling Landau prediction. However it ought to note the primarily paradoxical character of such experiments: the contradiction between the assumption of Maxwellian distribution function which is just setting as a result of collisions between electrons (and just with very strong dependence of the collision cross-section on electron velocity), and the assumption of collisionless plasma. There is shown particularly [7] that at the principal value IDI-sense collisionless damping (non-Landau) is possible for the low speed transverse waves (and presumably also for the longitudinal slow waves of analogous type) at accounting for distorting slight exceeding the distribution function above the Maxwellian distribution in the tail (with the further $|v_x| \simeq (1.5 \div 2)\sqrt{\overline{v_x^2}}$ replacing in integrands $v_x \to (v_x)_{eff}$).

Such experiments are extremely delicate and hard reproducible since they demand practically non-realizable conditions for the collisionless plasma with Maxwellian distribution function up to high velocity electrons, with ideal plasma isolation from the walls (despite the need to maintain the balance of the electron temperature $T_e$ of the gas-discharge plasma), plasma supporting with an external ionizing strong electrical field and additional distorting external longitudinal magnetic field, with insufficiently controlled or unknown plasma parameters. Stability of discharge column is supported with the energy balance at energy losses just due to electron collisions with molecules and the walls, and radiation.

Amplitude damping can also be a result of wave dispersion [8]. Besides that, experimental conditions can be affected by the non-Maxwellian (namely half-Maxwellian) distribution function near the plane of applying electrical field at $x=0$. A significant role can play present neutral particles. Thus the experimentally observed wave damping might be easily explainable otherwise than Landau damping.

Note again, that the solution $k_1, k_2 = F(\omega)$ with the wave functions of the type $\sim \exp(i\omega t \pm ikx)$, with the real dispersion equation and further usual transition to complex dispersion equation with $k \to k_1 - ik_2$, $k_2 \neq 0$ (see [3] and below *Appendix 1*) should not be obtained neglecting the imaginary part of dispersion equation.

We note in conclusion that owing to infinitely small additions in the precise quadratic in complex wave number $k$ dispersion equation [10], taking IDI in the sense of contour integrals can lead to possibility of very small imaginary additives to residua in poles $v_x = \pm \omega/k$ on the real axis which slightly shifts them to the complex plane. But this means that one should bypass around each pole not in the half-circle [2], but in the full circle (cf. [11]). At the same time, such shift of the singularity point would mean the contradictory possibility of integrating along the real axis without divergence of integrals, but with appearance their imaginary part.

Besides that in the theory [2] there remains unresolved the paradox of inevitable simultaneous presence of the waves with exponentially damping and growing in $-\infty < x < \infty$ amplitudes, so one has attracted the so called causality principle to acquit this in any way [12].

Nonlinear quadratic in perturbations terms of the wave equations in collisionless Maxwellian plasma lead to an appearance of non-damping overtones with multiple frequencies but not to the wave damping or growing at the given boundary field frequency [10], [13]. In this case the phases and amplitudes of overtones are defined by recurrent formulas with the phase velocity of the fundamental frequency wave [10], [14], ]17].

By definition, the condition of the total electron distribution function $f = (f_0 + f_1)$ being positive at all phases $f_1 \sim f_1(v_x)\cos(\omega t - kx + \varphi); \quad f_1(v_x) \sim (\partial f_0 / \partial v_x)/(\omega \pm k v_x)$ is correlated with the principal value sense of IDI, where near the poles velocity value $v_x$ is constrained by condition $|\omega/k \pm v_x| \leq \Delta v_x$ where $\Delta v_x$ may be defined by the cutting off condition $|f_1| \leq f_0$ correspondingly to the right and left bounds of this interval. It means that within this narrow interval kinetic equation loses sense and may be replaced with the physically reasonable finite transition between $f_1(v_x)$ equal $-|f_0(v_x)|$ and $|f_0(v_x)|$, both for the fundamental frequency wave and for overtones [10], [14], [15].

It ought to note that sometimes used qualitative considerations about acceleration of electrons with velocity more than the wave phase velocity can not be faithful since the wave phase velocity in plasma can be more than light velocity in vacuum. It implies apparently that one should account for the Cherenkov effect which can not be strictly described in the frame of classic physics and must be considered in the frame of relativistic theory with proper finding group velocity.

# *Appendix 1*
## Again on logical contradictions of Landau damping

V. N. Soshnikov[1]

Plasma Physics Dept.,
All-Russian Institute of Scientific and Technical Information
of the Russian Academy of Sciences
(VINITI, Usievitcha 20, 125315 Moscow, Russia)

### Abstract

Landau damping/growing at boundary condition of excitation $\sim \cos\omega t$ of a harmonic wave in collisionless ion-electron plasma contradicts to the law of energy conservation of a wave damping/growing in space. There is also no criterion of a choice either damping or growing solution in difference from always non-damping in the direction of their propagation Vlasov waves. Variety of other incongruities as consequence of Landau damping is specified also. When the collisional term in dispersion equation tends to zero, dispersion integral tends to integral in principal value sense with no reason for the occurrence of collisionless Landau damping/growing. Absence of explicit positivity and finiteness of the distribution function solutions of incomplete wave equations (without kinetic collisional terms) leads to need of imposing additional cutting off constraints with resulting positivity and finiteness of the full electron distribution function at the singularity point and finiteness of the dispersion integral. Cutting off the distribution function at the singularity point leads to real dispersion equation and unnecessary of using Landau rule of bypass the singularity point pole for determining spatial damping of steady over time electron waves. An artificial extension to complex wave number $k$ at real wave frequency $\omega$ leads to complex conjugate roots of the complex dispersion equation $k = k_1 - ik_2$ which violates the energy conservation law. Landau damping as a real physical phenomenon does not exist. A relation is established (see Appendices 2, 4) between the averaged over period real wave damping decrement $k_2$ and the added energy-exchange ("collisional") term of kinetic equation.

This Section is written in the frame of the conventional classic concept of collisionless damping of plasma waves according to complex dispersion equation obtained by arbitrary extrapolating the real wave number $k$ in the real dispersion equation to complex wave number $k \to k_1 - ik_2$.

According to for a long time standard concept, movement of a longitudinal harmonic electron wave in ion-electron-neutrals collisionless plasma is determined by the solution of a system of the kinetic equation and Maxwell equation of electrical field for distribution function $f = f_0(v) + f_1(v_x, x, t)$ where ones choose usually for function $f_0$ to be Maxwellian velocity distribution function and $f_1(v_x, x, t)$ is perturbation caused by electric field $E(x,t) \sim \exp[i(\omega t - kx)]$ at $x = 0$.

---
[1]E-mail: vikt3363@yandex.ru

Here arises singularity in the logarithmically divergent integral over longitudinal velocity $v_x$ with apparent inapplicability of such solution for obtaining dispersion equation. It is needed extraction of additional information from the basic equations outside the frame of the above approximation, in particular due to neglecting collision terms at solving kinetic equation.

The existing modification consists of artificial constraint of integrand of the indefinitely divergent integral (IDI) at which condition $|f_1| \ll f_0$ in all wave phases $(\omega t - kx)$ would be carried out at all velocities $|v_x|$ instead of non physical $f_1 \to \pm \infty$ near the singularity points $\pm \omega / k$ even at the finite dispersion integral.

This is very old problem of the dispute between A. Vlasov ([1], 1938) and L. Landau ([2], 1946) which has come to the end with the general acceptance of the so-called Landau rule of bypass the poles in complex plane $v_x$ leading to Landau damping, was discussed repeatedly in the work [3] and apparently, I think, this problem has not lost significance till now caused by very numerous in literature applications of Landau damping with enough doubtful experimental testing (collisionless Maxwellian electron plasma contradicts to possibility of neglecting electron collisions with each other, with molecules and the walls, not to mention exclusive difficulties of creating and supporting stationary homogeneous collisionless plasma).

It ought to note also the initially paradoxical character of such experiments: the contradiction between the assumption Maxwellian distribution function which is just setting as a result of electron collisions (and just with very strong dependence of the collision cross-section on electron velocity), and the assumption of collisionless plasma. A significant role can play present neutral particles. The thermal balance of electrons is supported by the account of balance of inflow of energy from an external field and energy outflow at collisions of electrons with walls (and may be, by radiation). One can measure only group velocity, but not the phase velocity. The role of usually used for plasma stabilization longitudinal magnetic and electrical fields is essential also. The real boundary conditions at the ends (transparent, reflecting, absorbing etc.) and in the middle of the plasma column with locally Maxwellian or half-Maxwellian distribution at $x = 0$ and on boundaries and so on, distribution function play the essential role. The approximation of infinite in $(y, z)$ plane waves (along direction $x$) is also problematic. These effects substantially mask very weak Landau damping/growing.

Collisionless Landau damping of monochromatic wave in the case of electron distribution function $f = f_0 + f_1$ at assuming finite $|f_1|$, $f_1 \sim f_1(v_x, \omega, k) \exp[i(\omega t - kx)]$, $k \equiv k_1 - ik_2$ and real $\omega$, propagating in rarefied electron-ion-neutrals plasma, is accepted as a cornerstone of contemporary physics of plasma with huge number of applications. It is based on Landau's *ad hoc* postulated proposal to eliminate logarithmic divergence of the dispersion integral over velocity $v_x$ in the dispersion equation $\omega = f(k)$ connecting variables $\omega$ and complex $k$, a rule of bypass around the pole of singularity points $v_x = \pm(\omega / k_1)$ in space of $v_x$ with a shift of integration contour upwards or downwards into the plane of complex velocities $v_x$ in analytical continuation of $f_1(v_x)$ that leads in general case either to complex values $\omega \to \omega \pm i(\delta\omega)$ or $k \equiv k_1 \pm ik_2$; $|\delta\omega| \ll \omega$, $|k_2| \ll |k_1|$. The more conservative and natural attempt is analogue of Van Kampen waves with arbitrary way cutting off function $f_1(v_x) \le f_{1\max}(v_x)$ on left and right of singularity point or choice integration ranges $\Delta v_x$ near to singularity points $v_x = \pm \omega / \operatorname{Re} k$ with integrating along the rest of real axis $v_x$ and the presence of modulated (entrained) by the wave an oscillating longitudinal beam of the small part of resonant electrons.

At $k_2 = 0$, the value $\delta\omega \ne 0$ defines decrement of damping (or increment of growing, depending on any choice of the integration contour of bypass around the singularity point) of the *running* wave $\sim \cos(\omega t - k_1 x) e^{\pm \delta\omega \cdot t}$ in time $t = x \cdot (k_1 / \omega)$. At $\delta\omega = 0$ corresponding with value $k_2 \ne 0$ it defines the wave collisionless damping (or growing) in space $\sim \cos(\omega t - k_1 x) e^{\pm k_2 x}$. In the case of boundary condition $\sim \cos \omega t$, $\delta\omega = 0$ we obtain damping (or growing) in space wave with $k_2 \ne 0$. These cases of solution of the complex dispersion equation with complex either $\omega$ or $k$ correspond for running wave to equality $(\delta\omega) \cdot t = k_2(\omega / k_1) \cdot t$ and in the intermediate case by the sum $[\delta\omega + k_2(\omega / k_1)] \cdot t$ (at $x = (\omega / k_1) \cdot t$). But damping/growing in space contradicts to the elementary law of energy conservation in absence of energy dissipation or transfer by collision interactions of electrons $f_1$ and $f_0$ or present molecules or any radiation losses with appropriate energy-exchange terms in kinetic equation (which problem does not exist however for Van Kampen non-damping Vlasov waves with $k_2 = 0$ modulated electron beam of the traveling electrons).

Landau damping can not be identified also with "evanescent energy" wave of total internal reflection with conservation of energy (no dissipation) [6]. It is necessary to notice also that both choices, exponentially damping or growing at propagation wave solutions, in Landau theory are mathematically absolutely equivalent. Thus, for example, if at $x=0$ perturbation $\sim \cos\omega t$ with real $\omega$ is applied, there is opened a question on behavior and sewing together solutions at $x<0$ and $x>0$ or the simultaneous presence of both waves at $x>0$.

Using for this purpose the so-called causality principle [10] is represented absolutely artificial and unreasonable. In general case of infinite plasma, possible steady over time solution with singularity point integrals in the principal value sense can be the non-damping Vlasov wave or standing wave (oscillations [5]) with energy conserving which is consistent with neglecting collision term in the kinetic equation. Account for inter collisions $\sim f_1 \cdot f_0$ and generally between all electrons and molecules leads to complex $k$ with arising dissipation damping in space, but steady over time oscillations and waves.

Besides it there is possible appearance of uncertainty in bypassing around the pole on a semicircle or on a $v_x$ full circle, considering that at presence even very weak collision damping, the pole is really always a little displaced at complex $k$ from the real axis $v_x$ (cf. also [4]). Arbitrarily small value $\text{Im}\,k$ corresponds to a pole in the complex plane $v_x$ with non zero residue at this pole. Conversely, this pole defines at its bypass the previous value $\text{Im}\,k$ without any possibility to distinct generally the collisional and collisionless parts of $\text{Im}\,k$, correspondingly the part of Landau damping, which also is not arbitrarily small. Note that Landau damping is due to addition to the principal value of the dispersion integral (Vlasov) half-residuum (or full residuum) in the pole $v_x = \omega/\text{Re}\,k$. Just a set of contradictions and ambiguities arises because of the seemingly excessive addition this pole residuum to the principal value of dispersion integral.

However it would be possible to consider various improved ways to provide convergence of dispersion integral, for example, to enter passing around the pole which is displaced from the real axis $v_x$ into the complex plane by a some predetermined distance, so that the total wave damping could self-consistently correspond to the estimated rate of an exchange of energy in dissipation collisions and Landau damping. In this case the way of passing around the pole on the full circle allows any way, unlike Landau damping (when the pole is on real axis $v_x$), to satisfy at least partially the law of energy conservation at presence of dissipation (instead of the account for the collision term directly in the kinetic equation) by a choice the value of the pole displacement from real axis. Neglecting small imaginary part, the damping solution can be sought presumably in real form

$$\text{Re}\,f_1(x,t,\omega,v_x) \sim e^{-x\,\text{Im}\,k}\left(\partial f_0/\partial v_x\right)\frac{(\omega - v_x\,\text{Re}\,k)\cdot\cos(\omega t - x\,\text{Re}\,k)}{(\omega - v_x\,\text{Re}\,k)^2 + (v_x\,\text{Im}\,k)^2} \quad \text{(using } E(x,t)\sim i e^{i(\omega t - x\,\text{Re}\,k + ix\,\text{Im}\,k)}\text{)} \quad (1)$$

which at $\text{Im}\,k = 0$ leads to not divergent dispersion integral, furthermore, $\text{Re}\,f_1 = 0$ at $v_x = \omega/\text{Re}\,k$. When $v_x \to \omega/\text{Re}\,k$ and $\text{Im}\,k \to 0$ the result depends on the choice of priority of their tending which is determined only by physics (non mathematics!) considerations and constraints. In this case the most robust choice appears to be principal value sense for dispersion integral without addition non zero complex residuum of the pole which creates $\text{Im}\,k$, with not smoothed transition through real axis $v_x$ at $+\text{Im}\,k \to -\text{Im}\,k$. Such a smoothed transition can be achieved only in the case of rejection the addition to the dispersion integral in sense of principal value the non zero complex residuum in the pole $v_x = \omega/\text{Re}\,k$ with Landau rule of change bypass directions when passing through the real axis $v_x$.

While $\text{Im}\,k$ in Eq. (1) determines the damping without distinction collisionless and collisional damping, it can be interpreted according to the energy conservation law, as determining collisional and collisionless energy loss. There can be arbitrary small collision correction to the function $f_1(v_x)$ with an arbitrary small shifting the singularity point $v_x = \pm\omega/k$ into the complex plane $v_x$. This shift may correspond to a damping of the wave due to the transfer of energy from electrons $f_1$ to electrons $f_0$ and molecules or the wave amplification due to the back energy transfer.

This leads to an evident total uncertainty of the acceptable mathematical versions of calculation divergent integral at: (1) along the real axis $v_x$ at an arbitrary small values $\text{Im}\,k \to \pm 0$ integration reduces to finding the integral in the principal value sense (Vlasov) and non damping waves; (2) along the real axis at an arbitrary small values $\text{Im}\, k \to \pm 0$, with a bypass around the singularity point on the real axis $v_x$ along the semicircle in the complex plane $v_x$ (Landau rule for collisionless damping); (3) along the real axis $v_x$ at an arbitrary small values $\text{Im}\, k \to 0$ with a bypass around the singularity point $v_x = \omega/k$, $\text{Im}\, k \to \pm 0$ on the real axis $v_x$ along the full circle in the complex plane $v_x$ (Alexeff and Rader, [4]). And in all cases $f_1$ is much more than $f_0$ near the singularity point with a finite value of the dispersion integral.

In the case of complex $k$ with diverse $\text{Im}\, k \equiv k_2 \neq 0$, for finding complex root of the complex dispersion equation one can use the method of analytical continuation of $f_1(v_x)$ into complex plane $v_x$ (Landau), but with arising symmetrical relative to the real axis $v_x$ poles corresponding to values $\pm k_2$ and taking their full (not half!) residua. Thus, if Landau damping exists, there is possible a variety of decrements of Landau damping depending only on $f_0(v)$ and $\omega$, but not depending on any energy exchange terms.

Since borders of integration interval near to the singularity point $v_x = \omega_0/k_0$ along the axis $v_x$ remain uncertain, the base simplest decision is taking the integral in the principal value sense. This Vlasov rule corresponds to the energy conservation law at real $k$. The additional external information defining a modulated beam of electrons which are entrained by the wave with its velocity $\sim \omega_0/k_0$ along axis $v_x$ can be also useful for any choice of integration borders $\Delta v_x$ (if non-symmetrically different) right and left of the singularity point. The follow-up adding the complex half-residuum in singularity point $v_x = \omega/\text{Re}\, k$ with a choice the direction of bypass around pole is usual Landau rule for collisionless damping with small $k_2$, but now with violation of the energy conservation law.

But at small $\text{Im}\, k$ cutting off $f_1$ can be necessary near to the singularity point according to the condition $|f_1| \leq f_0$ i.e. with natural physical constraint $0 < f_0 + f_1 \leq 2f_0$. Function $|f_1|$ can not be larger $f_0$ because then in opposite phase

$$f_1(v_x)\cos(\omega t_1 - kx_0 + \varphi) \to -f_1(v_x)\cos(\omega t_2 - kx_0 + \varphi), \quad \omega t_2 = \omega t_1 \pm \pi \qquad (2)$$

where $x_0$ is arbitrary selected coordinate, $f = f_0(v_x) + f_1(v_x, x, t)$ can become negative. Because of this, the electrons move in waves making local reverse movement ("capsize") almost without forming directed flow. Translational motion with the moving modulated electron beam is defined by the value of $\partial f_0(v)/\partial v_x$ in expression $\int f_1(v_x) v_x dv_x \neq 0$. Finiteness of dispersion integral can be reached also by the account of small collision damping with finite integrating then along the axis of real values $v_x$ needlessly to consider any poles of "collisionless damping" in the plane of complex values $v_x$. Here the energy balance should be reached with small non-zero collision term of the kinetic equation, which is usually neglected.

Translational modulated weak beam of electrons (Van Kampen wave) can occur probably always due to cutting off the distribution function between $\mp |f_1(v_{x1})|_{\max}$ and $\pm |f_1(v_{x2})|_{\max}$ (or vice versa) in a narrow region $\Delta v_{x1}$, $\Delta v_{x2}$ left and right near the point of singularity $v_x = \omega/k$ and the dependence $f_0$ on velocity component $v_x$. To find both values $f_{1\max}$ and dependence $f_1(v_x)$ in the range of $\Delta v_{x1}$, $\Delta v_{x2}$ it is necessary to use a qualified kinetic equation taking into account all the electron-electrons, electron-ions and electron-neutral molecules collisions, which is beyond the scope of this paper. But one can approximately assume that $f_1(v_x)$ in this range has the form of a smooth curved transition from $f_1(v_{x1}) = -f_0(v_{x1})$ to $f_1(v_{x2}) = f_0(v_{x2})$ (or, respectively, vice versa). At this case, the hypothetical possibility of emergence of a very weak modulated electron beam at entraining electrons by the main wave is determined by difference of the total number of electrons $\int f_1(v_x) dv_x$ with negative and positive $f_1(v_x)$ which increases with increasing steepness of $\partial f_0/\partial v_x$.

Therefore Landau's proposal (qualified as discovery) is based only on the subjective postulated procedure of finding the root of complex dispersion equation in approach of small $|k_2|$, but in this case being accepted and recognized by all scientists even despite occurrence of specified above incongruities and ambiguities. Note that in the today's survey and educational literature with description Landau damping, the simple theoretical derivation in the original [2], [7], [8], [9] is repeated almost without changes and some new proofs. At the same time Landau damping often is included as working part in difficult theoretical plasma constructions, for example, nonlinear Landau damping, standing waves with Landau damping, optical wave front reversal in collisionless plasma, plasma echo, in astrophysical applications both to longitudinal and transversal waves, numerous other applications with resonant electron beams, astrophysical applications, and so on.

Steady over time wave/oscillatory solutions can depend naturally on boundary conditions at $x \to \pm 0$, plasma slab sizes (infinite, half-infinite, finite), position of the point $x=0$ relative to possibly partially reflecting slab edges and so on, what allows to obtain differing steady over time solutions in $f_1(x,t)$ by varying conditions on slab edge boundaries or procedures in tending slab sizes to infinity, so it can be Vlasov $v_x \to \omega/k$ waves and oscillations.

The integrand in the part of complex $f_1$ in dispersion equation near the singularity point $v_x$ (here $v_x$ is longitudinal speed of electrons in $f_1$; $f_1 \sim (\partial f_0/\partial v_x)/(\omega - kv_x)$ can be written as

$$\frac{\partial f_0/\partial v_x}{(\omega - v\cdot\operatorname{Re}k) - i\cdot v\operatorname{Im}k} = \frac{(\partial f_0/\partial v_x)\left[(\omega - v\cdot\operatorname{Re}k)\right] + i\cdot\operatorname{Im}k\cdot v}{(\omega - v\operatorname{Re}k)^2 + (\operatorname{Im}k\cdot v)^2} = \frac{(\partial f_0/\partial v_x)\operatorname{Re}k\cdot\Delta v}{(\operatorname{Re}k\cdot\Delta v)^2 + (\operatorname{Im}k\cdot v)^2} + i\cdot\frac{(\partial f_0/\partial v_x)\operatorname{Im}k\cdot v}{(\operatorname{Re}k\cdot\Delta v)^2 + (\operatorname{Im}k\cdot v)^2}$$

(3)

where $v = v_0 \mp \Delta v$, and $(\Delta v)$ designates speeds $v$ in the range of bounds $\omega/\operatorname{Re}k \mp |\Delta v|$, defined by the limiting values $|f_1(v)|_{max}$ near $v_0 = \omega/\operatorname{Re}k$ with a smooth transition curve from $-|f_1|_{max}$ to $|f_1|_{max}$ (or vice versa) and with $f_1 \sim \pm 0$ in the middle part of the interval within the range of $\pm|\Delta v|$ ($|f_1(v)| < |f_1|_{max}$ in $|\Delta v|$, correspondingly $|\Delta v|$ right and left may be somewhat uncertain and depend also upon the term $\operatorname{Im}k < 0$). The waveform at the forward direction of wave movement: in front is pit, behind is hill, or vice versa, depending on the initial boundary (at $x=0$) phase of the electric field and periodically varies with the phase $(\omega t - x\operatorname{Re}k)$ of the wave.

Complex value of $f_1$ with $k_2 \neq 0$ leads to complex dispersion equation with two complex conjugate roots $k_1 + ik_2$ determining the wave number $k_1$ and decrement/increment $\mp k_2$ of the waves, corresponding to collisionless damping/growing of this plasma waves.

**Note 1.** In this study one implies that $|f_1|_{max}$ can reach amount $|f_1(v_x)|_{max} \leq f_0(v)$. For these values the wave equations are non-linear, with the emergence of the non-damping wave harmonics with multiple frequencies [5]. A qualitative justification of the made approximations with values $|f_1|$ up to $f_0$ may be a very small interval $|\Delta v_0|$ of velocities $v_x$ in which the distribution function $|f_1|$ has the maximum value $|f_1(v_x)|_{max} = f_0$ and nearly antisymmetry of $f_1(v_x)$ in this interval with very little effect on the dispersion relation.

## Conclusion

Accounting for naturally always present for any actual rarefied plasma the extremely weak non-zero collisional absorption parameter of electron wave $\operatorname{Im}k$, already as such, leads ultimately to a finite real dispersion integral in the principal value sense (Vlasov) and makes to be senseless literally understood instructions on calculation Landau damping (see [7]) with imaginary term of the pole half-residuum (or full residuum) in the complex plane $v_x$ (out of the real axis $v_x$) in the dispersion equation. However the real part of the dispersion equation also contains the imaginary part $\operatorname{Im}k$. Further refinement, quite natural and practically with very small numerical changes in solution of the dispersion relation $\operatorname{Re}k = f(\omega)$, has to keep the definition requirement of positivity and finiteness of the distribution function in the process of its approximate evaluation near singularity point $v_x \sim \omega/\operatorname{Re}k$ which is natural consequence of the partly actual presence of electron collisions or collisionless reverse electron movement ridge-trough in running wave despite neglecting weak collision term in the kinetic equation.

Analysis of the dispersion equation obtained using real exponentially damping wave functions presented in *Appendix 2* shows that at coincidence of the real solutions of the real dispersion equation and the real part of the complex dispersion equation with integrand (3) when $\operatorname{Im}k \equiv k_2 \neq 0$, even at low $|k_2|$, both solutions are radically different kind. But in contrast to standard "Landau damping" in the derivation of which is used smallness of $|k_2|$, in the case of the complete complex dispersion equation the decrement $k_2$ may in fact have also arbitrary large values depending only on $f_0(v)$ and $\omega$, and not depending on any energy loss terms. Thus it should be noted also that at solving of the complex dispersion equation with the component (3), the wave number $\operatorname{Re}k \equiv k_1$ depends on the decrement $\operatorname{Im}k \equiv k_2$ (they are interdependent).

The use of dispersion relations obtained in accordance with the formal Landau regulations leads to a number of logical contradictions.

**Note 2.** Arguments of previous versions are extended and supplemented without changing final results, also more detailed progress has been made in Appendices 2, 4 and 6).

# *Appendix 2*

## On the wave damping in weakly collisional plasma

V. N. Soshnikov[1]

Plasma Physics Dept.,
All-Russian Institute of Scientific and Technical Information
of the Russian Academy of Sciences
(VINITI, Usievitcha 20, 125315 Moscow, Russia)

### Abstract

For collisionless electron-ion-molecule plasma the dispersion equation with an exponentially decaying complex wave functions can be obtained by generalization of the real dispersion relation for non-damping wave functions with the replacement of the real wave number $k$ by a complex $k \to k_1 \pm ik_2$. It is common to approximate present in it the logarithmically divergent on the real axis of the speed $v_x$ integral in the sense of the principal value with the addition of semi-residue at the singularity point $v_x = \omega/k_1$ in the complex part of the dispersion equation which determines the collisionless Landau damping [2], [4]. The complex part of the dispersion equation leads to collisionless damping with violation of the energy conservation law, that can be characterized as the Vlasov-Landau damping. It is shown that in the case of the real dispersion equation, solution for exponentially damping wave function is radically different from the complex solution of the dispersion equation and is possible only as averaged in the wave period in the presence of energy-exchange ("collision") terms in the kinetic equation with the addition of the energy conservation equation to determine the wave number $k_1$ and the damping decrement $k_2$.

### 1. Introduction

Collisionless Landau damping is described in all textbooks on plasma physics as theoretically long ago predicted physical phenomenon of exponentially damping/growing electron waves in a weakly collisional plasma [1], [2]. Theoretical deriving is based on the substitution of solutions in the form of complex running waves into the field equation and the kinetic equation with finding complex roots of the resulting complex dispersion equation, for example, for a given real frequency, where $k(\omega) = k_1(\omega) + ik_2(\omega)$ and $k_1$ is wave number, $k_2$ is decrement/increment of the electron wave. In this case the wave damping/growing is not consistent with the law of energy conservation and leads to a number of other contradictions [3].

___________________________
[1] E-mail: vikt3363@yandex.ru

Substitution into the original equations the real waves of the form

$$\exp(-k_2 x)\left[\cos(\omega t - k_1 x)\right] \text{ and } \exp(-k_2 x)\left[\sin(\omega t - k_1 x)\right]$$

leads to real dispersion equation containing two parameters $k_1$ and $k_2$ to be determined and (at given $\omega$) the need to add the second equation which naturally is the equation of energy conservation with addition the energy-exchange ("collision") terms to the kinetic equation. It is shown that in the absence of such terms the collisionless damping, which corresponds to the solution $k_2 \neq 0$, does not exist.

## 2. Analysis of complex dispersion equation

The well known solution for the electron velocity distribution function when considering electron waves in collisionless plasma is sought in the form of addition to the Maxwell electron distribution function $f_0(v)$ the complex perturbation (with real $\omega$)

$$f_1(v_x, x, t) \sim f_1(v_x) \exp[i(\omega t - kx)]. \tag{1}$$

After its substitution to the collisionless kinetic equation

$$\frac{\partial f_1(v_x, x, t)}{\partial t} + v_x \frac{\partial f_1(v_x, x, t)}{\partial x} + \frac{e}{m_e} E(x,t) \frac{\partial f_0(v)}{\partial v_x} = 0 \tag{2}$$

we obtain

$$f_1(v_x, x, t) = -\frac{e}{m_e} \frac{(\partial f_0/\partial v_x) E(x,t)}{i(\omega - kx)} \tag{3}$$

where the complex electric field in Maxwell equation

$$\frac{\partial E(x,t)}{\partial x} = \frac{e}{m_e} \int f_1(v_x, x, t) dv_x \tag{4}$$

is assumed to be equal

$$E(x,t) = E_0 \cdot i \exp[i(\omega t - kx)], \tag{5}$$

with in general case $k = \mathrm{Re}\, k - i\, \mathrm{Im}\, k \equiv k_1 - ik_2$. The complex dispersion equation $k = f(\omega)$ contains logarithmically divergent integral as the real at $\mathrm{Im}\, k = 0$ expression

$$k \equiv \mathrm{Re}\, k - i\, \mathrm{Im}\, k = -\omega_0^2 \int \frac{(\partial f_0(v)/\partial v_x)}{\omega - k v_x} dv_x. \tag{6}$$

This leads to the relation between the actual observed real parameters of the real part of the dispersion equation $\omega$, $\mathrm{Re}\, k$, $\mathrm{Im}\, k$ where $\omega_0$ is Langmuir frequency $\omega_0^2 = 4\pi e^2/m_e$ and $\mathrm{Im}\, k > 0$ characterizes the wave damping $\sim \exp(-x\, \mathrm{Im}\, k)$

$$\mathrm{Re}\, k = -\omega_0^2 \int \frac{(\partial f_0(v)/\partial v_x)(\omega - v_x \mathrm{Re}\, k)}{(\omega - v_x \mathrm{Re}\, k)^2 + (v_x \mathrm{Im}\, k)^2} dv_x \tag{7}$$

with the real part of the electric field (5) in the plane $x = 0$

$$\operatorname{Re}E(x,t) = -E_0 e^{-x\operatorname{Im}k} \sin(\omega t - x\operatorname{Re}k). \qquad (8)$$

Virtual imaginary part of the dispersion equation

$$\operatorname{Im}k = \omega_0^2 \int \frac{(\partial f_0(v)/\partial v_x) v_x \operatorname{Im}k}{(\omega - v_x \operatorname{Re}k)^2 + (v_x \operatorname{Im}k)^2} dv_x \qquad (9)$$

leads, when $\operatorname{Im}k \to 0$, either to identity $0=0$ at accounting for the cutting off $f_1(v_x)$ condition

$$0 < f_0(v) + f_1(v_x) < 2f_0(v) \qquad (9a)$$

(with neglecting some violation of perturbation condition $|f_1| \ll f(v_0)$), or to the expression with divergent at $k_2 \to 0$ integral

$$1 = \int \frac{v_x (\partial f_0(v)/\partial v_x)}{(\omega - v_x \operatorname{Re}k)^2 + (\operatorname{Im}k)^2} dv_x. \qquad (10)$$

The latter result with the strange solution $\pm k_2 \neq 0$ ("Vlasov damping" (9) with $|+k_2| = |-k_2| \neq 0$) can be significantly modified by using the cutting off condition (9a) for $f_1(v_x)$ and is violating the energy conservation law, with impossibility of the smooth transition $\operatorname{Im}k \to 0$ which indicates the inadequacy of using nonlinearly complex expressions unrelated to the energy conservation law to describe the relationships between real physical quantities with a side effect of working with nonlinearly complex equations.

But at the same time to each given value $\omega$ there corresponds root of the complex dispersion equation with $\operatorname{Re}k$, $\operatorname{Im}k$, that is collisionless damping with $\operatorname{Im}k \neq 0$ while without having any need to add a residuum (or half-residuum) of any absent pole to the finite at $k_2 \neq 0$ integral (9) (as it is made for example in [4] at now usual deriving collisionless Landau damping using approximate method of analytical continuation of $f_1(v_x)$ into complex plane $v_x$).

In conclusion, it should be noted however that it is impossible to obtain the single relation between the one-valued physical observable real quantities based on the various identical records (various forms) of the same complex nonlinear dispersion equation with various versions of extraction from it the real and imaginary parts, but with the one-valued complex root.

Exact solution of the complex dispersion equation is the basis of collisionless Vlasov-Landau damping/growing ( = "Landau Damping") with an approximated $k_2$ using analytical continuation in complex plane $v_x$ with in fact its $\pm k_2$-mirror complex poles relative to the real axis $v_x$ and with their full, not half, residua) $f_1 \sim \exp(\pm k_2 x)$ with violation of the energy conservation law.

## 3. Analysis of real dispersion equation

With direct substitution of initially real values $\operatorname{Re}E(x,t)$ and $\operatorname{Re}f_1(v_x, \operatorname{Re}k, \operatorname{Im}k, x, t)$ (in which $\operatorname{Im}k \equiv k_2$ determines the wave damping $\sim \exp(-k_2 x)$) in the kinetic and Maxwell equations we obtain real dispersion equation in different form

$$1 = -\omega_0^2 \frac{ab}{k_1 a + k_2 b} \int \frac{(\partial f_0/\partial v_x)}{(\omega - k_1 v_x) b + k_2 v_x a} dv_x, \qquad (11)$$

where we use the notation

$$\operatorname{Re}k \equiv k_1, \quad \operatorname{Im}k \equiv k_2 > 0, \quad \cos(\omega t - k_1 x) \equiv a, \quad \sin x(\omega t - k_1 x) \equiv b. \qquad (12)$$

Since the dispersion equation contains two unknown variables $k_1$, $k_2$ to determine them one needs to add an independent energy conservation equation of collisional energy balance. However, the exact damping solution with independent of $x, t$ values $k_1(x,t)$, $k_2(x,t)$ even neglecting derivatives $\partial k_1/\partial x$, $\partial k_1/\partial t$, $\partial k_2/\partial x$, $\partial k_2/\partial t$ of the next order of smallness (e,g. even by using the some average values $k_1(x,t)$ and $k_2(x,t)$) do not exist in this somewhat artificial initial $|f_1(v_x,x,t)| < f_0(v)$ formulation of the problem for any $k_2 \neq 0$ due to integrand terms with $a$ and $b$.

This dispersion equation is analogous to equation (7) only if $k_2 \equiv \mathrm{Im}\,k = 0$. Its distinction from the equation (7) can be understood by comparing with the trivial elementary illustrative example of nonlinear inequality for arbitrary real values $c$, $d$, $g$, $h$:

$$cd \neq \mathrm{Re}\big[(c+ig)(d+ih)\big]. \qquad (13)$$

In the case of linearly complex expressions there is no "mixing" of the real and imaginary parts of a variety of complex variables, and equating to zero the real and imaginary parts of the complex dispersion equation should lead to the same real result $k = f(\omega)$. But note that the separation of the dispersion equation for the real and imaginary parts $\hat{D}F = F_r + iF_i = 0$ is changed by multiplying it by any complex number with conservation of the complex root. Besides that, when is a nonlinearly complex $F$, we have $\hat{D}\mathrm{Re}\,F \neq \mathrm{Re}\,\hat{D}F$.

Eq. (11) is similar in the form to the above expression **(7)** (but radically different at account for (9) which is divergent at $k_2 \to 0$). That is, even at cutting off the distribution function $f_1(v_x)$ in accordance with the condition of positivity of the total distribution function

$$|f_1(v_x,x,t)| < f_0(v) \qquad (14)$$

at $k_2 v_x a \to 0$ we obtain in (11) dispersion integral in the principal value sense with non-damping wave.

As a first step of the process of finding damping solution one can use some non zero value $k_2$, but then the collision term kinetic equation of the type

$$-A f_1(v_x)\exp(-k_2 x)\cos(\omega t - k_1 x) \qquad (15)$$

to be added to the right hand side of kinetic equation, with which the dispersion equation becomes

$$1 = \frac{-ab}{ak_1 + bk_2} \cdot \omega_0^2 \int \frac{\partial f_0(v)/\partial v_x}{(\omega - k_1 v_x)b + ak_2 v_x - A}\,dv_x. \qquad (16)$$

In the process of tending $A \to ak_2 v_x$, the dispersion integral is finite, which corresponds to its calculation in the principal value sense in analogy with the expression (7), or using cutting off the function $f_1(v_x)$.

Apparent internal contradiction is that in a hypothetical Landau damping with even without cutting off $f_1(x)$, the dispersion integral is finite, and in the absence of poles of the integrand (11) any need to add imaginary residue/half-residue to dispersion integral that leads to Landau damping [4] is completely absent.

In this simplified initial formulation of the problem (11) the dependences of $k_1$ and $k_2$ on $x,t$ can not be calculated (the problem is insoluble), but apparently possibility exists of their calculation in the approach of averages values.

The main result of the made analysis is proof that the so-called Landau damping does not exist as a real natural phenomenon.

It is possible, however, to obtain the averaged over oscillations values $k_1$ and $k_2$, at least approximately, for $|k_2|\lambda \lesssim 1$, where $\lambda$ is wave length $\lambda = 2\pi/k_1$, by setting $A = ak_2 v_x$ with

$$1 = \frac{-a}{ak_1 + bk_2} \int \frac{\partial f_0(v)/\partial v_x}{(\omega - k_1 v_x)}\,dv_x, \qquad (17)$$

then

$$f_1(v_x) = \frac{e}{m_e} \frac{\partial f_0(v)/\partial v_x}{(\omega - k_1 v_x)} E_0 \qquad (18)$$

where expression $a/(ak_1 + bk_2)$ is replaced by the value averaged over the oscillation period with improper integral in the sense of the principal value

$$\left(\frac{a}{ak_1+bk_2}\right)_{av} = \frac{1}{2\pi}\int_0^{2\pi}\frac{k_1-k_2\cdot(\tan y)}{k_1^2-(k_2\cdot\tan y)^2}dy = \frac{1}{2\pi}\int_0^{2\pi}\frac{k_1}{k_1^2-(k_2\cdot\tan y)^2} \quad . \tag{19}$$

Here $|k_2|\lambda$ must be small, since in general the average values $\overline{\exp(-|k_2x|)}$ and $\exp(\overline{-|k_2x|})$ can differ by orders of magnitude at large $|k_2|x$ (as it is, for example, in optics of atomic spectral lines with large peaks of $|k_2|x$).

Thus, to obtain a physically reasonable result it isn't enough to nullify the imaginary parts of the plasma parameters in the resulting dispersion equation. You must repeat the entire derivation of the dispersion equation, but using initially real equivalent wave functions with $k_2 \neq 0$.

Average values $k_1$ and $k_2$ should be determined at including besides (17), (19) the integral law of energy conservation, respectively, $k_2$ can be large enough in (7), (9) to satisfy condition (14), but it can be small, and then one should any way implement cutting off function $f_1(v_x)$ according to condition (14) which is practically necessary, unlike (7), in all cases, according to the expression (17).

Integral (17) can be calculated in the principal value sense that corresponds to the form of the integral (6) obtained using complex variables. Also integral (19) can be taken in the sense of principal value.

It is inconceivable that there would not be arbitrarily small wave damping in the presence of arbitrarily small but non-zero terms of collisional kinetic equation.

Thus, it is clear that we obtain by (16) or (17) at $\text{Im} k \to 0$ non-damping collisionless Vlasov solution [1] with the integral in the principal value sense; hence proposed by Landau [2] eight years later adding to complex Vlasov dispersion equation the imaginary residue at the pole $v_x = \omega/\text{Re} k$ [4] leading to the so-called Landau damping appears not correct. The real problem is to find the roots of the complex dispersion equation without any artificial residua additives.

Due to the smallness of the range $|\Delta v_x|$ near the singularity point $v_x \sim \omega/\text{Re} k$ and smallness $|\partial f_0(v)/\partial v_x|$ at large $v_x$, corrections due to the missing collisional term in the kinetic equation influence usually very weakly on the solution of the dispersion equations (16), (17), however obtaining solution for $k_2$ requires to attract additional new equation (since the dispersion equation contains at least two parameters $k_1$ and $k_2$ to be determined that takes into account the energy loss in collisions of electrons with each other and with atoms and molecules present in plasma. "Collisional" term can include any energy exchange processes of any nature.
In computing, for example, the modulated flux of electrons entrained by non-damping Van Kampen waves at $\text{Im} k = 0$ it is necessary to carry out a smooth cutoff of distribution function $|f_1(v_x)|$ near the critical point $v_x \sim \omega/\text{Re} k$ in accordance with the condition (14).

Integral energy balance can be written, for example, as relation of the type

$$\int f_1(v_x)k_2v_x\Delta\varepsilon(v_x)dv_x = \int f_1(v_x)\left[n_e\sigma_1(v_x)v_x\Delta\varepsilon_2(v_x)+n_m\sum_{n>1}\sigma_n(v_x)v_x\Delta\varepsilon_{n+1}(v_x)\right]dv_x \tag{20}$$

with possible resonances of $k_2$ in dependence on $\omega$ and arising excited molecules, where $\Delta\varepsilon_1(v_x)$ is energy loss per electron associated with a decrease in the number of electrons with an additional wave propagation velocity $v_x$; $\Delta\varepsilon_2(v_x), \Delta\varepsilon_3(v_x)$ ... , are portions of transmitted energy; $\sigma_n(v_x)$ are collision cross sections; $n_e$ is electron density; $n_m$ is density of atoms and molecules. In this

$$\Delta\varepsilon_1 \sim \frac{m_e}{2}\left[(v_{xM}+v_x)^2-v_{xM}^2\right] \tag{21}$$

where $v_{xM}$ is proper velocity of electron along the axis $x$ in Maxwell distribution $f_0(v)$ independent on the wave speed. To expression (20) one can add also any other actual energy exchange processes.

It is assumed that at extreme complexity of calculations using original kinetic equation, the proposed approximations with averaging $k_1, k_2$ and calculation averaged $\overline{k_2} = \text{const}$ based on the energy balance represent the main features of the damping plasma waves.

A more detailed exposition is presented in [3], where are market some obvious logical contradictions that arise when using Landau rules: violation of the energy conservation law in the damping/growing waves; the appearance of exponentially divergent solution at $|x| \to \infty$; uncertainties in calculation of the residue or the half residue near/on the real axis of the wave velocity $v_x$; jump transition between the non-zero values of the damping parameter $\pm \text{Im} k \to \mp \text{Im} k$ when choosing the bypass direction around the pole (cf. inspiring causality principle in [5]) or at infinitesimal displacement of the pole across the real axis $v_x$; contradictios of the calculating residue or half residue at the pole.

# 4. Conclusion

There is considered general ideology of solving the problem of wave propagation in a collisionless plasma with the phenomenological account of collisional damping parameter (decrement) $\text{Im} k \neq 0$ determined by collision term in the kinetic equation, with the substantiation of impossibility of collisionless Landau damping, which leads to a number of obvious logical contradictions [3].
Thereby:

1. Using technology of complex variables to solve the equations for waves $\sim \exp[i(\omega t - kx)]$ with complex $k \equiv k_1 - ik_2$, $k_2 \neq 0$ leads to results that are not equivalent to the solution of equation with real physically observable variables and waves

$$\sim \exp(-k_2 x) \times [\cos(\omega t - k_1 x) \text{ or } \sin(\omega t - k_1 x)]. \qquad (22)$$

Nonlinear complex expressions lead to errors in the transition to the real relations of physically observable real quantities with violation energy conservation law.

2. Limiting transition to the singularity point $v_x \to \omega/k_1$ in which the integrand of dispersion integral remains finite for any values $v_x$ arbitrarily close to $\omega/k_1$ of the type of $1/(\omega - k_1 v_x)$, leads to the dispersion integral in the principal value sense without the need to supplement with imaginary residue at the singularity point, including the case of the dispersion equation with complex $k$.

3. Moreover, the requirement of positivity of the total distribution function leads to the necessity of its cutting off near the singularity point with resulting the finite (constrained) integrand and finite dispersion integral.

4. When adding collision terms into the kinetic equation, surprisingly simple and natural averaging of dispersion equation over the period of oscillations leads to the dispersion equation for two unknown parameters $(k_1, k_2) = \text{const}$.

5. The second equation (20) required to determine the values $(k_1, k_2)$ is the second real averaged over period $\lambda = 2\pi/\omega$ energy balance equation, and the exponential damping of the wave at arbitrarily infinitesimal collision term leads to infinitesimal value $k_2$.

6. In the case of collisionless plasma such averaging is not possible, and the solution of the dispersion equation (11) with $k_2 \neq 0$ and exponentially damping wave with $(k_1, k_2 \neq 0)$ does not exist.

7. It is evident also that finding complex root of the complex dispersion equation and the proof of equivalence (or nonequivalence) of the collisionless "Vlasov damping" and collisionless "Landau damping" may represent only an abstract mathematical interest and have very slight indirect or no relation to the actual interconnection of physically observable real quantities $\omega, k_1, k_2$ while using the primary real variables with the real dispersion equation. Complex and real toots of the wave equations are disparate mathematical objects of different categories.

Thus, the main results are: (1) nonphysical negative electron distribution functions must be cut off, what leads to elimination of singular points in the dispersion equations; (2) the use of complex expressions for the wave functions with solving nonlinear complex dispersion equation leads to erroneous relations between physically observable real wave parameters and the emergence of collisionless damping (ghosts of complex roots of complex dispersion equations); (3) solving the real wave dispersion equation with constant parameters $k_1$ and $k_2 \neq 0$ (at constant real $\omega$) is possible only when averaged over the period oscillation values $k_1, k_2$. At the same time their determination requires addition of the second equation with $k_1, k_2$ which can be an energy balance equation when adding collision terms into kinetic equation.

8. In view of the above, for real $\omega, k$, the replace of integral in the sense of principal value in Vlasov work [1], which becomes an ordinary finite integral at cutting off the integrand function $f_1(v_x)$ at a singular point $v_x \sim \omega/k$, with a complex construction using Laplace transform, that leads to the appearing physically absurd simultaneous existence of two collisionless damping and growing plasma waves [2], becomes entirely unfounded and contradictory.

[This Appendix is correction of the author's comment pp. 515 – 516 of the Russian translation: Ф. Клеммоу, Дж. Доуэрти. Электродинамика частиц и плазмы. Изд. Мир, М., 1996, 518 с., [6].]

**Note 1.** English translation of the article being submitted in January 2014 to the Russian "Журнал Технической Физики" (ЖТФ) (Journal of Technical Physics, RF).

**Note 2.** This now updated *Appendix 2* (in Russian) was submitted to Russian-language Журнал Технической Физики (Journal of Technical Physics, RF).The article was rejected because of lack of novelty in February 2014 on the basis of the peer-review of known plasma physicist A. A. Rukhadze with including his text [*]. In this text he outlined the very interesting history of this problem, and pointed out that A. A.Vlasov even in 1945 spoke of a collisionless damping with no presence of any integration in the complex plane, which later after 1946 was requested and named as Landau damping, when Landau shifted the contour of integration in the dispersion integral over velocity $v_x$ to the complex plane with adding residuum in the singularity point, fitting it to produce the desired result. There are pointed out various further modifications including account for short-range and long-range forces, Coulomb cut off and other factors in formally collisionless plasma.

With the all correctness of this statement, however, in this connection I should indicate explicitly not denied specific results of my article: (1) the above given fact of contradictions of Landau damping is not considered, as we have; (2) predicted by Vlasov in 1945 collisionless damping can be result of using complex form of the nonlinear dispersion equation as noted above (see expression (9)), besides introduced in 1946 by Landau adding complex integration contour and residuum in the pole of the singularity point; (3) exponential damping/growing of electron waves is possible only when one adds appropriate collision terms in the right hand side of the kinetic equation; (4) the correct result can be obtained only with solution of the real form of dispersion equation, averaging over oscillations and adding the energy conservation equation; (5) finiteness of the dispersion integral is achieved with cutting off nonphysical negative values of the electron distribution function. Thus all singularities are eliminated. Is it not new? (however see p. 36 !).

These items are not discussed everyone in peer-review and qualified generally as already well known solved problems. However it is wrong way, in particular, to explain supposed collisionless Vlasov-Landau damping (with using Sokhotsky-Plegel theorem which establishes a connection between parameters $k_1$, $k_2$, $\omega$ of Landau damping and Vlasov damping) in application to the nonlinear complex dispersion equation, whereas it is appropriate to apply for this only the real dispersion equation with real physically observable values with in principle different in its sense result; further it is not commented the elementary necessity of very simple but crucial cutting off the negative electron distribution function and inseparable coupling the wave damping with collision terms of kinetic equation, furthermore, my denial (as the main result, contrary to Rukhadze) of this collisionless damping as a real physical phenomenon. Roots of the complex dispersion equation are not identical to the root of the real dispersion equation (different categories!).

*$^*$Рухадзе А. А.* // Vlasov A.A. and collisionless Landau damping. Институт общей физики им. А. М. Прохорова РАН, М., ул. Вавилова 38, С. 1-10. (Year unknown). See also: Рухадзе А. А. //Инженерная физика № 3, 2014 (in print). (References coming in this form personally from Rukhadze).

Carried out in my paper relatively very simple analysis leads to the conclusion that an analytical description of the damping/growing electron waves in plasma are impossible without adding appropriate collision terms in the collisionless kinetic equation.

The principal result of this paper is demonstration of the serious consequences of commonly used for many years in physics of wave processes in weak collisional plasma the techniques in the form of transformations of nonlinearly complex expressions that leads to completely erroneous conclusions about the true relations of real physical observables. In particular, values $k_1$, $k_2$ in Eq. (11) depend on $x$, $t$ unlike equations (7), (9).

Trivial explanation of many-years ambiguities of Vlasov-Landau wave damping in weakly collisional plasma by uncritical using results of transformations the complex nonlinear equations with complex wave functions is shocking.

In this case, there is the open question: whether now Landau damping is considered and can be considered as an outstanding discovery of plasma physics?

**Note 3.** Much of the material presented in *Appendix* 2 is now published in the article: *Soshnikov V. N.* On the wave damping in weakly collisional plasma. International Journal of Theoretical and Mathematical Physics, 2014, v.4, n.2, pp. 58 -62.

# *Appendix 3*
# Recent visiting the old problem of the wave damping in collisionless plasma

V. N. Soshnikov[1]

Plasma Physics Dept.,
All-Russian Institute of Scientific and Technical Information
of the Russian Academy of Sciences
(VINITI, Usievitcha 20, 125315 Moscow, Russia)

## Abstract

When using long-accepted classical approach to obtain the dispersion equation of the plane electron wave in collisionless plasma excited by applied in section (plane) $x=0$ electric field $E_0 \cos(\omega t + \varphi)$ the usual substitution in kinetic equation and Maxwell equation of the electric field the real expressions for damping running waves of the electric field and the perturbation electron distribution function with the constant wave number $k_1$ and decrement $k_2$ leads to dispersion equation, solutions of which $k_1 = F_1(\omega)$, $k_2 = F_2(\omega)$ are also functions of coordinates and time $x,t$ contradicting previously accepted condition $(k_1,k_2) = \text{const}$. Suitable reasonable solutions can be obtained only when $k_2 = 0$ or for averaged over small oscillation values $k_1$, $k_2$ with adding in the latter case to kinetic equation arbitrarily small collision terms together with the equation of energy conservation in collisions to determine averaged values $k_1$ and $k_2$.

## 1. Introduction

The dispersion equation of non-damping [1] or damping [2] electron waves in collisionless plasma arising under the influence of a periodic electric field in the plane section $x=0$, is obtained by substitution of expected solutions in the form of running waves into the kinetic equation

$$\frac{\partial f_1(v_x,x,t)}{\partial t} + v_x \frac{\partial f_1(v_x,x,t)}{\partial x} + \frac{e}{m_e} \frac{\partial f_0(v)}{\partial v_x} E(x,t) = 0 \qquad (1)$$

where $f_0(v)$ is electron Maxwell velocity distribution function, $f_1(v_x)$ is periodic perturbation of distribution function, and Maxwell equation of exiting electric field is

$$\frac{\partial E(x,t)}{\partial x} = 4\pi e \int_{-\infty}^{+\infty} f_1(v_x,x,t) dv_x. \qquad (2)$$

Substituting in the general case exponentially decaying solution in its real form

$$E(x,t) = -E_0 e^{-k_2 x} \sin(\omega t - k_1 x); \quad f_1(v_x,x,t) = f_1(v_x) e^{-k_2 x} \cos(\omega t - k_1 x) \qquad (3)$$

we obtain

---

[1]E-mail: vikt3363@yandex.ru

$$1 = -\omega_0^2 \frac{ab}{k_1 a + k_2 b} \int \frac{\partial f_0(v)/\partial v_x}{(\omega - k_1 v_x)b + k_2 v_x a} dv_x \qquad (4)$$

where $\omega_0$ is Langmuir frequency $\omega_0^2 = 4\pi e^2/m_e$ and we use the notation

$$\cos(\omega t - k_1 x) \equiv a, \quad \sin(\omega t - k_1 x) \equiv b \quad (k_2 > 0). \qquad (5)$$

## 2. Discussion

The exact damping solution with independent on $x,t$ values $k_1, k_2$ (even neglecting derivatives $\partial k_1/\partial t$, $\partial k_1/\partial x$, $\partial k_2/\partial t$, $\partial k_2/\partial x$ of the assumed next order smallness) does not exist in this somewhat artificial initial formulation of the problem for any $k_2 \neq 0$ due to integrand terms with $a$ and $b$. But it seems possible that there might exist damping solutions with continuous transition $k_2 \to 0$, while the value of the integrand and the dispersion integral remain finite at all $k_2 \neq 0$ the absence of any poles on the real axis $v_x$, defining the Landau damping in collisionless plasma in using transformations with presentations of functions and variables in complex form. However, note here, that such a continuous transition of solution between a certain value of the root $k_2 \neq 0$ and $k_2 = 0$ in the case of complex roots of complex dispersion equation does not exist.

It is quite obvious that the finite integral in (4) tends to its principal value when $k_2 \to 0$. Moreover, the condition of positivity of the total distribution function is

$$|f_1(v_x, x, t)| < f_0(v) \qquad (6)$$

at

$$f(v_x) \to \frac{e}{m_e} \frac{\partial f_0(v)/\partial v_x}{(\omega - k_1 v_x)} E_0. \qquad (7)$$

As a step of the process of finding damping solution one can use some non zero value $k_2$ but the collision term of kinetic equation of the type

$$-A f_1(v_x) \exp(-k_2 x) \cos(\omega t - k_1 x) \qquad (8)$$

should be added then into the right hand side of kinetic equation, and the dispersion equation becomes

$$1 = \frac{-ab}{ak_1 + bk_2} \cdot \omega_0^2 \int \frac{\partial f_0(v)/\partial v_x}{(\omega - k_1 v_x)b + ak_2 v_x - A} dv_x. \qquad (9)$$

In the subsequent process of tending $A \to ak_2 v_x$ the dispersion integral is finite, which corresponds to its calculation in the principal value sense, and we obtain

$$1 = \frac{-a}{ak_1 + bk_2} \cdot \omega_0^2 \int \frac{\partial f_0(v)/\partial v_{x(\;)}}{(\omega - k_1 v_x)} dv_x \qquad (10)$$

where expression $a/(ak_1 + bk_2)$ can be replaced by the value averaged over the oscillation period:

$$\left(\frac{a}{ak_1 + bk_2}\right)_{av} = \frac{1}{2\pi} \int_0^{2\pi} \frac{k_1 - k_2(\tan y)}{k_1^2 - (k_2 \tan y)^2} dy. \qquad (11)$$

However, this integral contains singular points and can not be calculated without some additional assumptions (as integral in the sense of principal value).

Average values $k_1$ and $k_2$ ($|k_2|$ must be less than $\lesssim k_1/2\pi$, since, in general, the average values $\overline{\exp(\pm k_2 x)}$ and $\exp(\overline{k_2 x})$ can differ by orders of magnitude at large $k_2$ as it is, for example, in optics of atomic spectral lines with large peaks of $k_2$) should be determined by the equations (10), (11) and the integral law of energy conservation which can be written, for example, as relation of the type

$$\int f_1(v_x)k_2 v_x \Delta\varepsilon(v_x)dv_x = \int f_1(v_x)\left[n_e\sigma_1(v_x)v_x\Delta\varepsilon_2(v_x)+n_m\sum_{n>1}\sigma_n(v_x)v_x\Delta\varepsilon_{n+1}(v_x)\right]dv_x \qquad (12)$$

with possible resonances of $k_2$ in dependence on $\omega$, where $\Delta\varepsilon_1(v_x)$ is energy loss per electron associated with a decrease in the number of electrons with an additional wave propagation velocity $v_x$; $\Delta\varepsilon_2(v_x)$, $\Delta\varepsilon_3(v_x)$ ... , are portions of transmitted energy; $\sigma_n(v_x)$ are collisions cross sections; $n_e$ is electron density; $n_m$ is density of atoms and molecules. At this

$$\Delta\varepsilon_1 \sim \frac{m_e}{2}\left[(v_{xM}+v_x)^2 - v_{xM}^2\right] \qquad (13)$$

where $v_{xM}$ is proper velocity of electron along the axis $x$ in Maxwell distribution $f_0(v)$. In expression (12) one can add also any other energy exchange processes.

A more detailed exposition is presented in [3], where are marked some obvious logical contradictions that arise when using Landau rules: violation of the energy conservation law in the damping waves; the appearance of exponentially divergent solution at $|x|\to\infty$; uncertainties in calculating the residue or the half residue near/on the real axis of the electron velocity $v_x$; jump transition between the non-zero values of the damping parameter $\pm k_2 \to \mp k_2$ when choosing the bypass around the pole (cf. inspiring causality principle in [4] ) or at infinitesimal displacement of the pole across the real axis $v_x$; contradictions of the calculating residue or half residue at the pole.

Inadequacy of Landau damping is also in using nonlinearly complex functions and variables in the nonlinear terms that leads to an inadequate transition to real relations of physically observable quantities. It is unclear how to extract uniquely the right relation between real physically observable parameters out of complicated nonlinear expressions with complex functions and quantities.

Simple calculations to obtain the real dispersion equation for the real exponentially damping/increasing harmonic wave functions (only *single* equation with **two** unknowns!) lead to the conclusion that a complex solution of the complex dispersion equation for the nonlinearly complex wave functions, which reduces to the solution (now of *two* equations with **two** unknowns for the real and imaginary parts) have completely different physical meaning of procedure, and the use of complex wave functions with complex dispersion equation leads to different disparate results and thus is a logical and finally mathematical error. It is shocking, that collisionless Landau damping/growing does not exist as a real physical phenomenon.

But this does not mean that all articles with the usually negligibly small additional terms of "Landau damping" are wrong, if there are used the correct energy balance equations, be it pair collisions of particles or "collisionless" energy exchange of the type of collective collisions, of by radiation, or any quantum-mechanical process, or any other way of energy exchange. In the case of the real dispersion equation with two unknowns there must be added the equation of energy conservation. In the second case, we obtain two ghosts solutions without the energy conservation.

## 3. Conclusion

Substitution of exponentially damping solution with decrement $k_2$ for perturbation of the electron distribution function $f_1(v_x)$ excited by the periodic electric field applied in the plane (section) $x=0$ given in the form of real functions and variables shows, that the integrand at small $k_2$ must be cut off near the singularity point $v_x \sim \omega/k_1$ according to the condition (6), i.e. the integrand and the dispersion integral are always finite.

Wave damping $k_2$ is determined, besides using the dispersion equation by addition collision terms into kinetic equation and by them related law of energy conservation in collisions of particles.

Solution of the nonlinear complex dispersion equation which determines the values $k_1$ and $k_2$ at the same time, is not equivalent to a solution of the real dispersion equation which contains both parameters $k_1$ and $k_2$, to determine which you should add the second real equation of the energy conservation law.

Carried out in this paper relatively very simple analysis leads to the conclusion that an analytical description of the damping/growing electron waves in plasma are impossible without adding appropriate collisional terms into the collisionless kinetic equation.

**Note.** Appendix 3 has been prepared primary as reduced material of the *Appendix 2* to be submitted to an English-language Journal.

# *Appendix 4*
# Collisionless plasma waves damping as physical phenomenon does not exist[1,2]

V. N. Soshnikov[3]

Plasma Physics Dept.,
All-Russian Institute of Scientific and Technical Information
of the Russian Academy of Sciences
(VINITI, Usievitcha 20, 125315 Moscow, Russia)

## Abstract

Complex wave number of longitudinal damping plasma (electron) waves $k = k_1 - ik_2$ (at a real frequency $\omega$) is typically found as a complex root of a complex dispersion equation obtained by substituting the complex wave functions into the linearized collisionless kinetic equation and Maxwell equation for the electric field. One root of the complex equation is $(k_1; k_2 = 0)$ but there are also two equivalent roots $(k_1; \pm k_2 \neq 0)$. Roots with $k_2 \neq 0$ determine collisionless exponential damping/growing Vlasov-Landau waves leading to a number of contradictions, including violation of the energy conservation law. Solution of the real dispersion equation obtained by substituting real damping wave functions using physically observable real quantities is the most reasonable, but leads to a single equation containing the wave number $k_1$ and either decrement (or increment) $k_2$. These solutions have nothing with the complex roots $(k_1; \pm k_2)$, furthermore, in addition for finding both parameters $k_1$ and $k_2$ one must add the second equation which is equation of energy conservation and is determined by additional ("collision") term in the right-hand side of the kinetic equation, including various energy exchange interactions. All integrals occurring in the dispersion equation are finite due to cutting off perturbation of the electron distribution function $f_1$ and can be approximately calculated in the principal value sense. It is shown that collisionless plasma waves damping (Landau damping) in plasma theory contains a logic error, thus it can not be real physical phenomena. There is derived the connection between wave damping and additional collision terms of kinetic equation and the averaging procedure, allowing us to obtain averaged constant values of the wave number and decrement.

PACS numbers: 52.25 Dg; 52.35 Fp.
Key words: collisionless plasma waves; Landau damping; Vlasov non-damping waves; collisionless electron wave damping; plasma waves dispersion equation.

## 1. Introduction

Landau damping represented as the discovery in 1946, is now a truism (see for instance [1] ), expounded in all textbooks on plasma physics, firmly established truth in plasma physics, beyond any doubt, and actively used in a huge number of plasma physics works. It is also seen as an extension of Vlasov (1938), who derived the dispersion equation for the non-damping longitudinal waves in collisionless plasma.

These works are reduced to derive the dispersion equation by substituting in the linearized kinetic equation (1) and the Maxwell equation for the electric field (2) the electron distribution function as a sum of background Maxwellian function $f_0(v)$ and perturbations $f_1(v, v_x, x, t)$, and $E(x, t)$ in the form of running harmonic waves with amplitudes $f_1(v, v_x) \equiv f_1(v_x)$ and $E_0$:

$$\frac{\partial f_1(v_x, x, t)}{\partial t} + v_x \frac{\partial f_1(v_x, x, t)}{\partial x} + \frac{e}{m_e} \frac{\partial f_0(v)}{\partial v_x} E(x, t) = 0 \qquad (1)$$

---

[1] V. N. Soshnikov. Arxiv.org/physics/0610220
[2] Nonlinear collisionless equation with arising non-damping overtones is considered in Arxiv.org/physics/0711.1321
[3] E-mail: vikt3363@yandex.ru

$$\frac{\partial E(x,t)}{\partial x} = 4\pi e \int_{-\infty}^{+\infty} f_1(v_x,x,t)dv_x. \quad (2)$$

with the obtained dispersion relation

$$k = -\omega_L^2 \int \frac{\partial f_0(v)/\partial v_x}{\omega - kv_x} dv_x. \quad (3)$$

where $e = -|e|$ is electron charge; $\omega_L = (4\pi e^2/m_e)^{1/2}$ is Langnuir frequency; $m_e$ is electron mass.

## 2. Dispersion equations

The same equation in which now $k = k_1 - ik_2$, can be obtained by substituting to (1), (2) complex damping waves

$$E(x,t) = E_0 i e^{i(\omega t - kx)}; \quad \text{Re}[E(x,t)] = -E_0 e^{-k_2 x} \sin(\omega t - k_1 x). \quad (4)$$

Obviously, the complex dispersion equation now can be written as

$$k \equiv k_1 - ik_2 = -\omega_L^2 \int \frac{(\partial f_0(v)/\partial v_x)}{\omega - kv_x} dv_x = \quad (5)$$

$$= -\omega_L^2 \int \frac{(\partial f_0(v)/\partial v_x)(\omega - v_x k_1)}{(\omega - v_x k_1)^2 + (v_x k_2)^2} dv_x + i\omega_L^2 \int \frac{(\partial f_0(v)/\partial v_x) v_x k_2}{(\omega - v_x k_1)^2 + (v_x k_2)^2} dv_x. \quad (6)$$

The root $(k_1, k_2)$ of complex dispersion equation is determined by a joint solving two equations:

$$k_1 = -\omega_L^2 \int \frac{(\partial f_0(v)/\partial v_x)(\omega - v_x k_1)}{(\omega - v_x k_1)^2 + (v_x k_2)^2} dv_x \quad (7)$$

and

$$k_2 = -\omega_L^2 \int \frac{(\partial f_0(v)/\partial v_x) v_x k_2}{(\omega - v_x k_1)^2 + (v_x k_2)^2} dv_x. \quad (8)$$

Equation (8) becomes a trivial identity $0 = 0$ at $k_2 = 0$ what corresponds to the solution in the form of non damping waves. However, there are also two fully equivalent solutions $\pm k_2 \neq 0$ corresponding to collisionless damping, which should be discarded because of violation the energy conservation law and due to a number of other explicit contradictions (see [2], [3] ). Thus, $k_2 \neq 0$ is a side root which is not connected with any physical process. We can say that these solutions correspond to collisionless damping of Vlasov – Landau (see [2]). Complete their inconsistency also follows from the further consideration. Equivalence of the damping resulting from the solution of equation (8) with the ubiquitous Landau solution can have only an abstract mathematical interest.

With direct substitution of initially real values $\text{Re}\,E(x,t) = -E_0 b e^{-ik_2 x}$ and $\text{Re}\,f_1(v_x,x,t) = a e^{-k_2 x} f_1(v_x)$ (in which $k_2$ determines the wave damping $\sim \exp(-k_2 x)$ ), in the kinetic and Maxwell equations we obtain real dispersion equation in fully different form

$$1 = -\omega_L^2 \frac{ab}{k_1 a + k_2 b} \int \frac{\partial f_0(v)/\partial v_x}{(\omega - k_1 v_x) b + k_2 v_x a} dv_x \quad (9)$$

where $a \equiv \cos(\omega t - k_1 x)$, $b \equiv \sin(\omega t - k_1 x)$.

These collisionless solutions have nothing to do with the above designated complex roots of the complex dispersion equation (6), (7), (8). Furthermore, solutions of Eq. (9) with $k_1, k_2 \neq 0$ that would not depend on $x,t$ do not exist. However, one can try to obtain some averaged values $k_1, k_2 \neq 0$ over the wave period introducing the collision term $S$ on the right side of the kinetic equation with the dispersion equation of the form

$$1 = \frac{-ab}{ak_1 + bk_2} \cdot \omega_0^2 \int \frac{\partial f_0(v)/\partial v_x}{(\omega - k_1 v_x)b + ak_2 v_x - A} dv_x \quad (10)$$

taking

$$A \equiv ak_2 v_x; \quad S \equiv A f_1(v_x) e^{-k_2 x}. \quad (11)$$

In this case dispersion relation takes the form [2]:

$$1 = \frac{-a}{ak_1 + bk_2} \omega_0^2 \int \frac{\partial f_0(v)/\partial v_x}{(\omega - k_1 v_x)} dv_x; \quad (12)$$

$$\left(\frac{a}{ak_1 + bk_2}\right)_{av} = \frac{1}{2\pi} \int_0^{2\pi} \frac{dy}{k_1 + k_2 \tan y} = \frac{1}{2\pi} \int_0^{2\pi} \frac{k_1 dy}{k_1^2 - (k_2 \tan y)^2}; \quad (13)$$

$$f_1(v_x) = \frac{e}{m_e} \frac{\partial f_0(v)/\partial v_x}{(\omega - k_1 v_x)} E_0 \quad (14)$$

where all integrals can be taken in the principal value sense. Besides that, due to singularity of the function $f_1(v_x)$ at the point $v_x = \omega/k_1$ in Eq.(14) near which the kinetic equation does not apply, it is necessary cutting off $f_1(v_x)$ near this point in accordance with the condition of positivity the total distribution function $f = f_0(v) + f_1(v_x, x, t)$:

$$|f_1(v_x, x, t)| < f_0(v). \quad (15)$$

The required for finding averaged $k_1$ and $k_2$ the second integral energy conservation equation can be written, for example, as relation of the type

$$\int f_1(v_x) k_2 v_x \Delta \varepsilon(v_x) dv_x = \int f_1(v_x) \left[ n_e \sigma_1(v_x) v_x \Delta \varepsilon_2(v_x) + n_m \sum_{n>1} \sigma_n(v_x) v_x \Delta \varepsilon_{n+1}(v_x) \right] dv_x \quad (16)$$

(with further integration over $v_x$) and with possible resonances of $k_2$ in dependence on $\omega$ and arising excited molecules, where $\Delta \varepsilon_1(v_x)$ is energy loss per electron associated with a decrease in the number of electrons with an additional wave propagation velocity $v_x$; $\Delta \varepsilon_2(v_x)$, $\Delta \varepsilon_3(v_x)$ ..., are portions of transmitted energy; $\sigma_n(v_x)$ are collision cross sections; $n_e$ is electron density; $n_m$ is density of atoms and molecules. In this expression

$$\Delta \varepsilon_1 \sim \frac{m_e}{2} \left[ (v_{xM} + v_x)^2 - v_{xM}^2 \right] \quad (17)$$

where $v_{xM}$ is proper velocity of electron along the axis $x$ in Maxwell distribution $f_0(v)$ independent on the wave speed. To expression (16) one can add also any other actual arbitrary energy exchange process terms.

## 3. Conclusion

The above considerations mean complete difference from the long ago established approach in plasma physics to the derivation of dispersion equations.

Trivial explanation of many-years ambiguities of Vlasov – Landau wave damping in weakly collisional plasma by uncritical using results of right transformations of the complex nonlinear equations is shocking. There is the open question: whether now Landau damping is considered and can be considered as an outstanding discovery in plasma physics?

# Appendix 5
# Replay: On a Common Logical Error in Calculation and Applying the Complex Conductivities of Collisionless Plasmas

V. N. Soshnikov[1]

Plasma Physics Dept.,
All-Russian Institute of Scientific and Technical Information
of the Russian Academy of Sciences
(VINITI, Usievitcha 20, 125315 Moscow, Russia)

## Abstract

Damping/growing of electron waves in collisionless plasmas (kinetic equation with zero energy-exchange collision terms) does not exist contrary to the established among plasma physicists more than half a century ago the view about theoretical discovery of its existence as a natural and experimentally confirmed physical phenomenon.

## 1. Commentary

The recent discussion [1] – [4] (with not principal amendment in improper integral (18) in [2] ) clearly confirms the trivial logical (and ultimately mathematical) cause of widespread (see e.g. textbooks [5] - [7] and numerous in literature other articles and textbooks on plasma physics) erroneous description of non-existing collisionless damping of waves in both non-magnetoactive and in general case magnetoactive collisionless (in the sense of neglecting the collision energy-exchange terms in kinetic equation) plasmas. The reason is that all the quantitative laws of the nature are characterized as relations between the so-called physically observable detectable values (PhDVs), which in the case of their complex representation (before all further non-linear complex transformations!) have to be substituted into the real initial wave equations in the form of real combinations of complex conjugate sets as it is done, for example, in the form of direct substitution of sought PhDVs into the wave equations in the works [1] - [4] with the final results also in the natural form of relations of real PhDVs. Moreover, a direct unambiguous relation is derived between the wave damping/growing parameters (if present) with the defined collision terms of kinetic equation.

It is wrong, for example, using generalization to damping/growing plasma waves by moving on to a complex wave number $k \to k_1 - ik_2$ in the intermediate steps or the final result (in the form of the dispersion equation) in the case of initially real $k$, because in complex nonlinear equations mixing occurs of terms with $k_1$ and $k_2$ in the imaginary and real parts of the resulting complex equations and expressions, and it becomes impossible to extract out of them the true relations between PhDVs. For example, the multiplication of complex dispersion equation $F(\omega, k_1, k_2) = 0$ by imaginary $i$ or any complex value reverses or changes the real and imaginary parts of the complex $iF$ without changing complex roots of the equation. In these cases, there arises also discrepancy of PhDV real roots $k_1, k_2$ in only one single equation with adding second equation of energy conservation, and two equations for complex root of the complex dispersion equation or a variety of side terms of equivalent complex expressions with different forms of separation to imaginary and real parts at neglecting imaginary part of equation ("single equations for the ghosts real roots" also with adding the second equation of energy conservation). Thus, it is absolutely wrong detection of the true relations of PhDVs by separating the real part of the final complex dispersion equation and dropping the imaginary part, where the last leads to fictitious collisionless damping.
Solution of the complex dispersion equation determines collisionless damping/growing of fictitious complex waves, but not the actual real PhDV waves corresponding to different individual nature of dispersion equations for complex and real waves.

A typical case is non linear with respect to combinations of $k_1$ and $k_2$ resulting dispersion equation with complex wave functions in the initially linear wave equations. And it is due to initial complex nonlinearity between combinations of the real and imaginary values. Complex nonlinearity arises, for example, already in the simplest case of derivative $d\left(e^{i(\omega t - kx)}\right)/dx = -ik e^{i(\omega t - kx)}$ at complex $k$ with to it corresponding inequalities at extending $k \to k_1 - ik_2$ as an arbitrary but typical illustrative simple example of possible mixing real and imaginary parts in final real expressions (cf, also complex conjugate sums (14) with nonlinear mixing real and imaginary

---
[1] E-mail: vikt3363@yandex.ru

terms)

$$\mathrm{Re}\left(\frac{d}{dx}\exp[i(\omega t-kx)]\right)=-\mathrm{Re}\{(ik_1+k_2)\exp(-k_2x)[\cos(\omega t-k_1x)+i\sin(\omega t-k_1x)]\}= \qquad (1)$$
$$-\exp(-k_2x)[k_2\cos(\omega t-k_1x)-k_1\sin(\omega t-k_1x)]\neq \mathrm{Re}(-ik)\times\mathrm{Re}\{\exp[i(\omega t-kx)]\},$$

i.e. there are inadmissible anyway generalizations in the transition from real to complex $c$ and $d$ of the type $\mathrm{Re}(cd)\to\mathrm{Re}c\times\mathrm{Re}d$ (cf. also below Eq. (13)).

It is rather unclear how after transformations with complex wave functions for the case of usual complex wave functions with the damping $\sim e^{-ik_2x}$:

$$f_1(v_x,x,t)=f_1 e^{i(\omega t-kx)};\ E(x,t)=iE_0 e^{i(\omega t-kx)};\ k\equiv k_1-ik_2 \qquad (2)$$

unlike initial substitution real form:

$$\mathrm{Re}\,f_1(v_x,x,t)=f_1(v_x)\cdot e^{-k_2x}\cos(\omega t-k_1x), \qquad (3)$$
$$\mathrm{Re}\,E(x,t)=-E_0\sin(\omega t-k_1x);\ a\equiv\cos(\omega t-k_1,x);\ b\equiv\sin(\omega t-k_1x) \qquad (4)$$

where $f_1(v_x,x,t)$ is perturbation of the background electron Maxwellian function $f_0(v)$; $E(x,t)$ is perturbation wave of the external field $-E_0(\omega t;\ x=0)$.

Substituting real expression (3) and (4) into the real wave equations (kinetic and Maxwell equations)

$$\frac{\partial f_1(v_x,x,t)}{\partial t}+v_x\frac{\partial f_1(v_x,x,t)}{\partial x}+\frac{e}{m_e}\frac{\partial f_0(v)}{\partial v_x}E(x,t)=0 \qquad (5)$$

$$\frac{\partial E(x,t)}{\partial x}=4\pi e\int_{-\infty}^{+\infty}f_1(v_x,x,t)dv_x. \qquad (6)$$

where $e=-|e|$ is electron charge, $m_e$ is electron mass, and $v,v_e$ are electron velocities, leads to the real dispersion relation [1], [2]:

$$1=\frac{-ab}{ak_1+bk_2}\cdot\omega_0^2\int\frac{\partial f_0(v)/\partial v_x}{(\omega-k_1v_x)b+ak_2v_x}dv_x \qquad (7)$$

where $a\equiv\cos(\omega t-k_1,x)$, $b\equiv\sin(\omega t-k_1x)$, $\omega_0$ is Langmuir frequency $\omega_0^2=4\pi e^2/m_e$.

It means that hypothetical solutions of Eq. (7) with $k_1;k_2\neq 0$ that would not depend on $x,t$ do not exist. However, one can try to obtain some average values $k_1;\ k_2\neq 0$ over the wave period introducing the collision term $S$ on the right hand side of the kinetic equation with the dispersion equation of the form

$$1=\frac{-ab}{ak_1+bk_2}\cdot\omega_0^2\int\frac{\partial f_0(v)/\partial v_x}{(\omega-k_1v_x)b+ak_2v_x-A}dv_x \qquad (8)$$

taking

$$A\equiv ak_2v_x;\ S\equiv Af_1(v_x)e^{-k_2x}. \qquad (9)$$

In this case the dispersion equation takes the form [1], [2]:

$$1=\frac{-a}{ak_1+bk_2}\int\frac{\partial f_0(v)/\partial v_x}{(\omega-k_1v_x)}dv_x;\ \left(\frac{a}{ak_1+bk_2}\right)_{av}=\frac{1}{2\pi}\int_0^{2\pi}\frac{dy}{k_1+k_2\cdot\tan y} \qquad (10)$$

with the need of correction of Eq. (18) in [2] and

$$f_1(v_x)=\frac{e}{m_e}\frac{\partial f_0(v)/\partial v_x}{(\omega-k_1v_x)}E_0. \qquad (11)$$

where all integrals can be taken, including improper integrals (10), in the principal value sense. Besides that, due to singularity of the function $f_1(v_x)$ at the point $v_x = \omega/k_1$ in Eq. (11) near which the kinetic equation does not apply, it is necessary cutting off $f_1(v_x)$ in (11) nearby this point in accordance with the condition of positivity the total distribution function $f = f_0(v) + f_1(v_x)$:

$$|f_1(v_x, x, t)| < f_0(v). \qquad (12)$$

Options of cutting off the function $f_1(v_x)$ which are caused by a variety of any additional considerations determine features of the apparently always present, weak Van Kampen modulated electron beams.

The obtained results are in principle different from those obtained when ubiquitous, without any exceptions, using the dispersion equations in the case $k_2 \neq 0$ after substitution therein complex wave functions (2) (see [1]). In this case, the complex roots of the dispersion equation contain an artifact when appearing imaginary part due to mixed, real and imaginary terms with $k_1$ and $k_2$ entails respectively two equations for the real and imaginary parts of dispersion equation that does not correspond in no way to the law of energy conservation or any other physical law. In contrast, the real dispersion equation is single equation containing both parameters $k_1$ and $k_2$, while the second equation of the system for to find $k_1$ and $k_2$ is directly the new, second equation of energy conservation that independently binds the sought parameters $k_1$ and $k_2$ included in collisional energy-exchange terms $A$ and $S$ according to (9) (see, for example, [1], [2], where also are pointed out numerous contradictios of the so-called collisionless damping in the current formulation of the problem) and regardless of the interpretation of the nature and sense of the collision process. From the above, it also follows that the commonly used expressions for complex tensors of electrical conductivity and dielectric permittivity of collisionless plasma should be used directly only for the real $\omega$ and $\vec{k}$ or with substitution of initially complex conjugate expressions for PhVDs which significantly reduces the advantages of the using complex variables.

Use of complex wave functions can lead to complex dispersion equations with complex non-physical roots. The analytical mathematical formalism of complex tensor conductivity/permeability is inapplicable in the presence of any significant energy exchange processes, the more that the real damping waves have not the form of exponentially damping strongly sinusoidal propagating waves according to (7).

In general, the solution for PhDVs can be obtained by using the complex conductivity tensor $\sigma_{ij}(\omega, \vec{k})$ with $\omega > 0$ and complex $\vec{k}$ as a real solution of the dispersion equation of the form

$$\left[\vec{j}_i(\omega, \vec{k}) e^{i(\omega t - \vec{k}\vec{r})}\right] + [\text{compl. conjug.}] = \left[\sigma_{ij}(\omega, \vec{k}) E_j(\omega, \vec{k}) e^{i(\omega t - \vec{k}\vec{r})}\right] + [\text{compl. conjug.}] \qquad (13)$$

which should be followed by integration over $\omega, \vec{k}$, where $j_{\vec{v}_i}(\omega, \vec{k})$, $E_j(\omega, \vec{k})$ are complex components of integral Fourier correspondingly for electrical current and tension of electrical field. In this approach, the consideration of imaginary part of such form of dispersion equation which should lead to the collisionless damping becomes meaningless.

Thus, in general, the real dispersion equation and its solution may contain the angular coordinates of the real and imaginary parts of the initial wave function with $\cos\theta$ and $\sin\theta$, $\theta \equiv \omega t - \vec{r} \cdot \text{Re}\vec{k}$ which makes however impossible to use the solution of dispersion equation as it follows from equation (7).

In the derivation of Eq. (13) it was assumed solution in the form of Fourier transform (particularly, for example, presented in [7])

$$F(\vec{r}, t) = \int C(\omega) e^{-i\vec{k}\vec{r}} \cdot e^{i\omega t} d\omega \qquad (14)$$

with one-valued dependence $\vec{k} = f(\omega)$ at general case of complex $\vec{k}$. But in general, solutions have form of the double (in $\omega$ and $\vec{k}$) Fourier transform

$$F(\vec{r}, t) = \iint_\infty C(\omega, \vec{k}) e^{i(\omega t - \vec{k}\vec{r})} d\omega d\vec{k} \qquad (15)$$

with real $\omega, \vec{k}$, and such one-valued dependence $\vec{k} = f(\omega)$ can be not existing, i.e. to each $\omega$ there corresponds the whole spectrum of values $\vec{k}$, and solutions of the form $\sim e^{i(\omega t - \vec{k}\vec{r})}$ of the type (13) with complex $\vec{k}$ do not exist at all (as evidenced by the above example of the dispersion equation (7)). Also very relation (13) becomes invalid.

Moreover, theoretical contradictions are accompanied with poorly compatible conditions of proposed experiments on collisionless damping [1].

## 2. Conclusion

Using a two-parametric $k_1$, $k_2$ non physical complex root of the complex dispersion equation for the damping complex plasma waves in a collisionless plasma instead of detecting two real observable parameters of the real dispersion equation with adding two-parametric real energy conservation equation is common fundamental logical and, by inference, mathematical error available in the literature on Plasma Physics, which can lead to an erroneous conclusion about the existence of collisionless damping of plasma waves with derived non physical decrement.

Procedure (13) of the transition to real dispersion equation complemented by integration over *v* entirely changes the normal procedure for the application of the complex conductivity tensor of collisionless and collisional plasma and can lead to in principle different from existing dispersion equations and their solutions with collisionless damping.

Exact exponentially damping sinusoidal plasma wave solutions for both collisionless and collisional plasma do not exist.

Damping/growing of fictitious complex plasma waves with usual using complex conductivity and complex permittivity has nothing to do with damping/growing of real plasma waves characterized by physical observable quantities.

---

**Note:** Material of this *Appendix* was published in: International Journal of Theoretical and Mathematical Physics, **4**(4), pp. 156 - 158, 2014.

## *Appendix 6*

## Collisionless damping of plasma waves as a real physical phenomenon does not exist

V. N. Soshnikov[1]
Plasma Physics Dept.,
All-Russian Institute of Scientific and Technical Information
of the Russian Academy of Science
(VINITI, Usievitcha 20, 125315 Moscow, Russia)

**Abstract**

Trivial logic of collisionless plasma waves is reduced usually to using ***nonlinearly complex*** exponentially damping/growing wave functions to obtain a ***complex*** dispersion equation for their wave number $k_1$ and the decrement/increment $k_2$ (for a given real frequency $\omega$ and complex wave number $k \equiv k_1 - ik_2$), whose solutions are ghosts $k_1$, $k_2$ which do not have anything to do at $k_2 \neq 0$ with the solution of the *real* dispersion equation for the initial exponentially damping/growing *real* plasma waves with the physically observable quantities $k_1$, $k_2$ for which finding should be added, in this case, the second equation of the energy conservation law. Using a *complex* dispersion equation for the simultaneous determination of $k_1$ and $k_2$ violates the law of energy conservation, leads to a number of contradictions, is logical error, and finally also is the mathematical error leading to both erroneous statement on the possible existence of exponentially damping/growing virtual *complexly nonlinear* waves of collisionless plasma which is wrongly attributed to the actual *real* plasma waves. A brief discussion is also on formalism of complex conductivity and dielectric permittivity with the ability to use their real values.

---

[1] E-mail: vikt3363@yandex.ru

*PACS numbers*: 52.25 Dg; 52.35 Fp.

*Key words*: collisionless plasma waves; Landau damping; Vlasov non-damping waves; collisionless electron wave damping; dispersion equation of plasma waves.

# 1. Introduction

Dispersion properties of the *real* and *nonlinear complex* wave functions are not comparable, because are properties of physically observable and abstract mathematical objects of entirely different nature.

Simple calculations to obtain the real dispersion equation for the real exponentially damping/increasing harmonic wave functions (only *single* equation with *two* unknowns!) lead to the conclusion that a complex solution of the complex dispersion equation for the *nonlinear complex* wave functions, which reduces to the solution (now of *two* equations with *two* unknowns for the real and imaginary parts) have completely different physical meaning of procedure, and the use of *nonlinear complex* wave functions with complex dispersion equation leads to different disparate results and thus is a logical and finally mathematical error. It is shocking, that the well known collisionless Landau damping/growing does not exist as a real physical phenomenon.

But this does not mean that all articles with the usually negligibly small additional terms of "Landau damping" are wrong, if there are used the correct energy balance equations, be it pair collisions of particles or "collisionless" energy exchange of the type of collective collisions, of by radiation, or any quantum-mechanical process, or any other way of energy exchange. In the case of the real dispersion equation with two unknowns there must be added the equation of energy conservation. In the second case, we obtain two ghosts solutions without the energy conservation.

## 2. Relation of averaged over period exponentially damping/growing plasma waves with the collision term of kinetic equation

It is assumed that the longitudinal in direction $x$ electron wave functions, both real and complex, should satisfy the same equations: (1) the linearized collisionless kinetic equation

$$\frac{\partial f_1(v_x,x,t)}{\partial t} + v_x \frac{\partial f_1(v_x,x,t)}{\partial x} + \frac{e}{m_e} \frac{\partial f_0(v)}{\partial v_x} E(x,t) = 0 \qquad (1)$$

and Maxwell equation relating the electric field $E(x,t)$ and the charge density

$$\frac{\partial E(x,t)}{\partial x} = 4\pi e \int_{-\infty}^{+\infty} f_1(v_x,x,t) dv_x, \qquad (2)$$

and as a standard approach adopted for the full distribution function

$$f = f_0(v) + f_1(v_x,x,t); \quad |f_1| \ll f_0. \qquad (3)$$

where $e \equiv -|e|$ is electron charge, $m_e$ is electron mass, $\omega$ is real frequency, and $v$, $v_e$ are electron velocities. From physical considerations, it is assumed that at the points where the condition (3) may be violated it is necessary to carry out cutting off the function $|f_1|$.

In the case of testing complex wave functions, ones use as solution the expression

$$f_1 = f_1(v_x) e^{i(\omega t - kx)}; \quad k = k_1 - ik_2 \qquad (4)$$

and necessary condition for the existence of solution

$$E(x,t) = iE_0 e^{i(\omega t - kx)}. \qquad (5)$$

Similarly, in the case of real wave functions to receive a real tested solution ones must use the solution in the form

$$f_1 = f_1(v_x) e^{-k_2 x} a, \quad a \equiv \cos(\omega t - k_1 x); \quad E(x,t) = -E_0 e^{-k_2 x} b, \quad b \equiv \sin(\omega t - k_1 x). \qquad (6)$$

Both cases are discussed in [1], so the further results are given only for the case of physically observable real functions and variables.

Substitution of the real expressions (6) into the real wave equations (1) and (2) leads to the real dispersion relation [1]

$$1 = \frac{-ab}{ak_1 + bk_2} \cdot \omega_0^2 \int \frac{\partial f_0(v)/\partial v_x}{(\omega - k_1 v_x)b + ak_2 v_x} dv_x \qquad (7)$$

where $\omega_0$ is Langmuir frequency $\omega_0^2 = 4\pi e^2/m_e$.

It means that hypothetical solutions of Eq. (7) with constant $k_1; k_2 \neq 0$ that would not depend on $x, t$ do not exist. However, one can try to obtain some average values $k_1; k_2 \neq 0$ over the wave period introducing the collision term $S$ on the right hand side of the kinetic equation with the dispersion equation of the form

$$1 = \frac{-ab}{ak_1 + bk_2} \cdot \omega_0^2 \int \frac{\partial f_0(v)/\partial v_x}{(\omega - k_1 v_x)b + ak_2 v_x - A} dv_x \qquad (8)$$

taking

$$A \equiv ak_2 v_x; \quad S \equiv A f_1(v_x) e^{-k_2 x}. \qquad (9)$$

In this case the dispersion equation takes the form [1], [2]:

$$1 = \frac{-a}{ak_1 + bk_2} \int \frac{\partial f_0(v)/\partial v_x}{(\omega - k_1 v_x)} dv_x; \qquad (10)$$

$$\left(\frac{a}{ak_1 + bk_2}\right)_{av} = \frac{1}{2\pi} \int_0^{2\pi} \frac{dy}{k_1 + k_2 \cdot \tan y} \qquad (11)$$

with the need of correction of Eq. (18) in [2] and

$$f_1(v_x) = \frac{e}{m_e} \frac{\partial f_0(v)/\partial v_x}{(\omega - k_1 v_x)} E_0. \qquad (12)$$

where all integrals can be taken, including improper integrals (10), in the principal value sense. Besides that, due to singularity of the function $f_1(v_x)$ at the point $v_x = \omega/k_1$ in Eq. (11) near which the kinetic equation does not apply, it is necessary cutting off $f_1(v_x)$ in (11) nearby this point in accordance with the condition of positivity the total distribution function $f$ in (3)

$$|f_1(v_x, x, t)| < f_0(v). \qquad (13)$$

The required for finding averaged $k_1$ and $k_2$ the second integral energy conservation equation can be written, for example, as relation of the type

$$\int f_1(v_x) k_2 v_x \Delta \varepsilon(v_x) dv_x = \int f_1(v_x) \left[ n_e \sigma_1(v_x) v_x \Delta \varepsilon_2(v_x) + n_m \sum_{n>1} \sigma_n(v_x) v_x \Delta \varepsilon_{n+1}(v_x) \right] dv_x \qquad (14)$$

(with further integration over $v_x$) and with possible resonances of $k_2$ in dependence on $\omega$ and arising excited molecules, where $\Delta \varepsilon_1(v_x)$ is energy loss per electron associated with a decrease in the number of electrons with an additional wave propagation velocity $v_x$; $\Delta \varepsilon_2(v_x)$, $\Delta \varepsilon_3(v_x)$ ..., are portions of transmitted energy; $\sigma_n(v_x)$ are collision cross sections; $n_e$ is electron density; $n_m$ is density of atoms and molecules. In this expression

$$\Delta \varepsilon_1 \sim \frac{m_e}{2} \left[ (v_{xM} + v_x)^2 - v_{xM}^2 \right] \qquad (15)$$

where $v_{xM}$ is proper velocity of electron along the axis $x$ in Maxwell distribution $f_0(v)$ independent on the wave speed. To expression (14) one can add also any other actual arbitrary energy exchange process terms.

Note that the complex dispersion equation with the substitution of complex wave functions (3), (4) into (1), (2) at enough large $k_2 \neq 0$ has nothing to do with the real dispersion equation of the real Eq. (7) and leads to collisionless damping in violation of the law of energy conservation [1].

## 3. Inadequacy of the direct application of the concepts of complex electric conductivity/dielectric permittivity and other complex parameters

The laws of nature are determined by relations between the physically observable quantities, but in the case of complex values with their mathematical transformations, besides simple linear relationships, to extract the true relationships between physical observable quantities may be impossible, since the complex non-linearity arises even in the simplest cases, for example, in differentiating $d\exp(ikx)/dx$, as at real and all the more complex $k$.

This is confirmed by results obtained above with completely different meaning of the dispersion equations for the real and non-linear complex wave functions, with the "discovery" of the collisionless damping of the *real* plasma waves, which is attributed them from entirely mathematically correct collisionless damping of *non-linear complex* waves.

This raises the question of obtaining the true relationship between real physically observed values from usually used quantities such as the complex conductivity, which is not reducible to the simple use of its real or imaginary parts. For example, in the case of the complex tensor conductivity $\sigma_{ij}(\omega,\vec{k})$ (see [2]), to obtain the real current density $\vec{j}_i(\omega,k)$ without some improper mixing the real and imaginary parts of constituent complex terms foremost in calculation of complex $\sigma_{ij}(\omega,\vec{k})$, one must use real part in expressions of the type (16) because $\vec{j}(\vec{r},t)$ is real physically observable quantity in all possible more general complex combinations relating to

$$\text{Re}\int_{-\infty}^{\infty} \vec{j}_i(\omega,\vec{k})e^{i(\omega t-\vec{k}\vec{r})}d\omega = \left[\frac{1}{2}\int_{-\infty}^{\infty}\sigma_{ij}(\omega,\vec{k})E_j(\omega,\vec{k})e^{i(\omega t-\vec{k}\vec{r})}d\omega\right] + [\text{compl. conjug.}] \quad (16)$$

with the integration $\sigma_{ij}(\omega,\vec{k})E_j(\omega,\vec{k})\exp[i(\omega t-\vec{k}\vec{r})]$ over $\omega$ in accordance with the Fourier transform at $0\leq t\leq\infty$ and inverse Fourier transform at $-\infty<\omega<\infty$ with analytical continuation to complex $\omega$ in (16) in the lower and upper complex half–plane $\omega$ [6].

For real $\vec{j}(\vec{r},t)$ and its Fourier transforms with real $\omega$ using relation (16) results in identically the same result with and without complex conjugation. The difference arises only in the case of the Fourier transforms of complex $\vec{j}(\vec{r},t)$ that may occur, as shown in Sec. 2, at using "wild" non-linear wave functions with complex $\omega$ in the function of the complex conductivity. This leads to a collisionless damping of plasma waves or to elicitation the real part with neglecting small imaginary part of the type discussed in *Appendix 1*, pp 6, 8 in [5] which also leads to partly erroneous values $k_1, k_2$ in the expression $k=k_1-ik_2$.

*Collisionless damping is a consequence of this particular local error of "wild" non-linear extrapolated complex wave functions, but in fact, the general linear theory of complex plasma conductivity and complex dielectric permittivity is correct and when using Fourier transforms of boundary and initial conditions with non-damping boundary electric field, as a result of this are only non-damping complex or real wave functions.*

As it follows from Sec. 2, exact damping solution can exist only as an integrated set of real damping waves $\omega, k$ with different dispersion relations. It requires an entirely new approach to the calculation of the complex conductivity and dielectric permittivity in the presence of collisional energy exchange and to the calculation of real wave functions.

In the case of the component of Fourier expansion $\sim e^{ikx}$ with real $k$ ($k_2=0$ with the single root $k=f(\omega)$ of dispersion equation (7)), the dispersion equation coincides with the real dispersion equation for the real wave functions, moreover, it contains only one parameter to be finding: the wave number $k$ of non-damping waves. Thus the need for the energy conservation equation is eliminated (non-damping waves).

If the direct and inverse Fourier transforms are possible, both forms of writing dispersion relations in the input variables $\vec{r}, t$, as well as in variables $\omega, \vec{k}$ in correspondingly direct and inverse Fourier transforms must be real in order to avoid the appearance of two-parameter complex roots. Reality of both equations is the criterion of correctness of carried out mathematical transformations. However, there may be severe error, in addition to using the dispersion equation for the complex wave functions of the type (4), (5), when the approximate equality of the product of the Fourier transforms of the real functions may differ significantly from the product of the Fourier transforms of these functions in violation of the generally accepted theory of the complex conductivity and dielectric permittivity.

It is necessary to exclude the possibility of relationships with complex $\sigma(\omega,\vec{k})$ that are specific for using "wild" wave functions and lead to erroneous relations between physically observable parameters of the problem of the type discussed above in Sec. 2. Eq. (16) is equivalent to using only real part of the complex dispersion equation.

The wave damping corresponds to the Fourier transform in which for each real frequency $\omega$ corresponds the spectrum of waves with different real $\vec{k}$'s. This can significantly hinder or prevent the application of the theory of complex electrical conductivity and permittivity with the need to consider the totality of waves with different dispersion dependences $k=f_n(\omega)$. This is clearly demonstrated by Eq. (10) in Sec. 2, when the collisional kinetic equation with fixed $\omega$ is followed by a set of dispersion equations in (10) with a variety of $k_1$ in the terms $a$, $b$ and equation of energy conservation.

In connection with the above it is of interest to test the received default analyticity of the integrand $F(v_x) \equiv (df_0/dv_x)/(\omega - kv_x)$ in the plane of complex $v_x$ at an arbitrary transition from real $k$ to complex $k$ which requires the check the analyticity of $F(v_x)$ with Cauchy-Riemann criterion.

Simple replacement of the real wave number $k$ or frequency $\omega$ in Fourier transforms by complex $k$ or $\omega$ is logical unacceptable mistake that leads to non-physical effects such as collisionless damping.

4. **Expansion of the wave solutions in the case of the collisionless kinetic equation with a nonlinear term, which is due to the perturbation of distribution function, in overtones**

When substituting the electron distribution function in the form $f = f_0(v) + f_1(v_x, x, t)$ in the kinetic equation of longitudinal waves in a collisionless plasma (1) it is usually neglected nonlinear term of second order of smallness

$$\frac{|e|E_x(x,t)}{m_e} \cdot \frac{\partial f_1}{\partial v_x} \qquad (17)$$

in the right hand side of the kinetic equation. It may seem that this "collision" term will result in a loss of waves energy, respectively, to their damping. However, attempts were made to obtain non-damping solutions in the form of expansions in the overtones of the form

$$f_1(\vec{v}, v_x, v_z, x, t) = \sum_{n=1} F_n(\vec{v}, v_x, v_z) \cos[n(\omega t - kx + \varphi_1)], \qquad (18)$$

$$E(x,t) = \sum_{n=1} E_n(x,t) \cos[n(\omega t - kx + \varphi_1) - \pi/2]; \qquad (19)$$

with recurrent relations between $F_n$ and preceding parameters $f_n$, $E_n$ with indices $\leq n$.

Amplitudes $E_n$, $F_n$ are proportional to $(E_1)^n$, what provides the convergence of expansion at any enough small $E_1$, or more precisely, convergence parameter

$$\eta \lesssim \left| \frac{e}{m_e} \frac{E_1}{\sqrt{v_x^2}} \frac{\pi}{\omega} \right| \ll 1. \qquad (20)$$

It should be noted that the rather cumbersome transformations given in [3] were made without the use of complex variables and are brought to $n \leq 3$. All details can also be found in *Appendix 6* of the easily accessible work [5]. Calculations are made for both the longitudinal and transverse plasma waves, although in the process of calculation it should be made successive cutting off, according to the condition of positivity the function $f$, that involves some uncertainty in the result.

An interesting fact would be the experimental discovery of multiple overtones, that does not seem an insurmountable task. Note also, that due to the weak asymmetry of $f_0(v)$ near the singularity point $v_x = \omega/k_1$ (with cutting off $f_1(v_x)$), there should always exist weak modulated Van Kampen electron beams.

## 5. Polarization estimation hypothesis

The distribution function of charged particles in the plasma is characterized by the collective distribution of particle swarm on the coordinates and velocities in the created by them (or external) electric field, neglecting the individual interactions of each particle with its nearest neighbors, therefore, as a measure of the applicability of the collisionless approximation one can take comparison impact force $eE(x,t)$ of electric field $E(x,t)$ on the particle and the Coulomb force of interaction between two nearest particles. Conventionally, one can assume that the collisional interaction can be neglected, starting with the value $|eE(x,t)| \gtrsim e^2/r_{av}^2$ where $r_{av} \sim n_e^{-1/3}$ is average distance between charged particles in plasma, i.e. actual interacting of $E(x,t)$ with the charged particles is equivalent to the replacing

$$E(x,t) \rightarrow E_{eff}(x,t) = \gamma E(x,t), \quad \gamma \sim E_0/|e|n_e^{2/3} \sim \text{const}; \quad 0 < \gamma \lesssim 1; \text{ if } \gamma > 1, \text{ then } \gamma \sim 1 \qquad (21)$$

where $E_0$ is amplitude of the $E(x,t)$ and $n_e$ is average electron density.

The polarization hypothesis allows to estimate the parameters of electron waves in collisional plasmas with $\gamma < 1$, and can be a useful tool for the study of plasma waves.

The polarization hypothesis underlies the estimates of the parameters in the dipole-dynamic model of ball lightning (DDM BL) in [4].

## 6. Conclusion

There is proposed the fundamentally new approach to basic concepts of electron waves in a low-temperature plasma including:

- discussion of the possible existence of exponentially damping/growing sinusoidal waves in collisionless and weakly collisional plasma, non-existence of damping/growing waves in collisionless plasma [1];
- inadmissibility of using complex dispersion equations obtained for the specific non-linear complex wave functions with complex $\omega, \vec{k}$ instead of predominantly the real dispersion equations obtained for the real wave functions [1] or using real $\omega, \vec{k}$;
- non-linearity of kinetic equations due to the perturbation of the electron distribution function $f_0(v)$ by perturbation $f_1(v_x, x, t)$ in collisionless plasma, leads to the appearance of multiple (integer $n$, $\sin[n(\omega t - kx)]$, $\cos[n(\omega t - kx)]$) overtones without showing any effects of increasing-damping of waves [3];
- polarization hypothesis with the possibility of simple estimates with transport the parameters of plasma waves in weakly collisional plasma to a more collisional plasma by introducing such a parameter as the effective electric field strength $E_{eff}(x,t)$ [4].

The polarization hypothesis was actively used when creating a dipole dynamic model of ball lightning [4], but it was entirely distinct from the parameter estimates of the plasma waves with $\gamma = \text{const}$ and variable $E(x,t)$, wherever in DDM BL the constant was the atmospheric electric field $E_{env}$ with variable $\gamma$.

Dispersion properties of actual real and virtual non-linear complex wave functions are not comparable, because are properties of physically observable and abstract mathematical objects of entirely different nature.

The foregoing results are presented in their totality and discussed in detail in the cumulative work [5].

**Note:** This Appendix is accepted to publication in the next issue of the Journal of Advances in Physics (JAP, (**6**)(3), December 2014, p. 1291 – 1296).

______________________________________________________________________________________________

## SUMMARY

The laws of nature are characterized as relations between the physically observable real quantities. Therefore, in some cases, the use of physical real variables can be regarded as testing and probing of calculations of physical processes made by using surprisingly fruitful huge mathematical apparatus of the theory of functions of complex variables.

In the case of real exponentially damping plasma waves in collisionless plasma with a given frequency, dispersion equation contains two parameters to be determined: the real wave number and the damping decrement, to find these you need to add the second equation, which can be the equation of energy conservation.

In the case of complex exponentially damping wave functions with complex wave number $k = k_1 - ik_2$ which causes nonlinear complexity of wave functions, complex dispersion equation is equivalent to the two equations correspondingly for the real and imaginary parts, determining the two parameters: the real wave number $k_1$, decrement/increment $k_2$, and two unphysically, unnaturally simultaneously coexisting exponentially divergent and damping solutions $k_1 \pm ik_2$ and besides more obviously violating the law of energy conservation, regardless of any other equations.

In the case of zero-decrement (i.e. the linear complexity approximation) complex wave functions lead to the real dispersion equation to be determined with the real wave number $k_1$ at $k_2 = 0$ (non-damping waves).

Analysis of the real dispersion equation for physically observable real values of the wave number and the decrement for the real wave functions being the real part of the complex wave functions shows that damping wave solution $k_1$, $k_2$ can exist only in the presence of terms of the collisional kinetic equation of a certain type, which corresponds to the energy conservation law with the inclusion of energy-exchange terms of very general form. More precisely, the exact solution still does not exist here, but it makes sense of the solution averaged over the period of oscillations.

It is carried out extended consideration of the problem of exponentially damping/growing real and complex plasma waves starting from a number of logical contradictions of collisionless complex nonlinear plasma damping/growing waves and divergence of electron distribution function at the singularity points.

There is shown the logical fallacy of simple nonlinear extensions of linear complex expressions for the physical processes in the form of nonlinear complexity replacing the real values (for example, either a real wave number or frequency in complex wave) by a complex wave number or complex frequency in the widespread paradigm of damping electron waves in collisionless plasma.

All this leads to the need to revise the widely represented for more than half a century works, including textbooks, the problem of collisionless damping of plasma waves and the use of basic plasma physics functional parameters such as the complex conductivity and the complex permittivity in the case of nonlinear complex extensions in complex expressions for the wave functions. Only real dispersion equation can be considered correct because it does not lead to the collisionless damping. Collisionless damping in the sense of Landau damping does not exist as a physical phenomenon. At the same time there remains an open problem of the possible use of expansions in the complex or real wave functions for obtaining the real dispersion equation for collisional plasma.

As seen from above, the existence of collisionless damping/growing of plasma waves is determined with shocking trivial simplicity by erroneous use of nonlinearly complex expressions, which are obtained by arbitrary replacing the real values in the complex wave functions by complex variables.

* * *

# Inability of exponentially damping/growing plasma waves in collisionless plasma*

V. N. Soshnikov[1,2]

Plasma Physics Dept.,
All-Russian Institute of Scientific and Technical Information
of the Russian Academy of Science
(VINITI, Usievitcha 20, 125315 Moscow, Russia)

## Abstract

It is clear that the quantitative laws of nature are relations between physically observable quantities. Meanwhile, as a rule, the dispersion relation for the real plasma wave functions is identified with the dispersion relation for the complex wave functions of entirely different mathematical category. Exploring the exponential damping/growing of electron waves in collisionless plasma I came to the very strange, but in fact very simple and logical dilemma. For a given wave frequency $\omega$ of the real damping wave, the latter is characterized by two parameters: the wave number $k_1$ and decrement/increment $k_2$. Collisionless kinetic equation and Maxwell field equation, which relates the local electric field tension and electron charge, lead to the real dispersion equation $\omega = F(k_1, k_2)$, i.e. to single equation with two unknowns, to determine which you must add a second independent equation connecting them, which, of course, can be naturally the law of energy conservation.

## 1. Complex and real plasma waves in collisionless plasma

As a rule, ones obtain the dispersion equation by substituting into the original kinetic equation and the equation of the electric field the wave functions of complex form $\exp[i(\omega t - kx)]$, where $k \equiv k_1 - ik_2$. At $k_2 = 0$, we obtain the real dispersion equation which coincides with the dispersion equation in the case of real nondamping waves. But when $k_2 \neq 0$ we get a complex dispersion equation with complex roots, giving simultaneous solutions $k_1$ and $k_2$ which are not related to the energy conservation law or any other equations. And in this case, the parameter $k_2$ defines the so-called collisionless damping of electron (plasma) waves. ***This is fundamentally different result!*** Moreover, the above works lead to the conclusion that the exponentially damping real electron waves can be described only as waves with averaged over oscillation period the values $k_1, k_2$, and exist only when added to the kinetic equation some energy-exchange (collisional) terms of specific type.

The use of complex nonlinear exponentially damping/growing wave functions leads to a number of contradictions of collisionless damping/growing plasma waves and to far-reaching logical conclusions about the fallacy of commonly used extrapolating $k \to k_1 - ik_2$ of the complex wave functions $\exp[i(\omega t - kx)]$ including also it in Fourier transforms with real $k$ at the analysis of damping/growing plasma waves.

## 2. Real plasma waves in collisionless and collisional plasma

Consider the possibility of solving the dispersion equation in the case of exponentially damping real wave functions in the longitudinal electron wave with the electric field of the form

_______________________________________________

[1] Krasnodarskaya str. 51-2-168, 109559 Moscow, Russia.
[2] E-mail: v.n.soshnikov@gmail.com.

$$E(\omega,x,t)=-E_0 b\exp(-k_2 x t), \quad b\equiv\sin(\omega t-k_1 x), \qquad (1)$$

and perturbed electron distribution function of the form

$$f(v,v_x,x,t)\simeq f_0(v)+f_1(v_x)e^{-k_2 x}a, \quad a\equiv\cos(\omega t-k_1 x), \qquad (2)$$

where $f_0(v)$ is Maxwell velocity distribution, and due to

$$0<f_0(v)+f_1(v_x)e^{-k_2 x}a<2f_0(v), \qquad (3)$$

in singular points $|f_1(v_x)|\to\infty$ it is required cutting off $f_1(v_x)$ from physical considerations. Substituting $f(v,v_x,x,t)$ and $E(\omega,x,t)$ from Eqs. (1), (2) into collisionless kinetic equation

$$\frac{\partial f_1(v_x,x,t)}{\partial t}+v_x\frac{\partial f_1(v_x,x,t)}{\partial x}+\frac{e}{m_e}\frac{\partial f_0(v)}{\partial v_x}E(x,t)=0 \qquad (4)$$

and Maxwell field equation

$$\frac{\partial E(x,t)}{\partial x}=4\pi e\int_{-\infty}^{+\infty}f_1(v_x,x,t)dv_x, \qquad (5)$$

one obtains

$$f_1(v_x)=-\frac{e}{m_e}E_0\frac{\partial f_0(v)/\partial x}{(\omega-k_1 x)b+k_2 v_x a} \qquad (6)$$

and dispersion equation

$$1=\frac{-ab}{ak_1+bk_2}\cdot\omega_0^2\int_{-\infty}^{\infty}\frac{\partial f_0(v)/\partial v_x}{(\omega-k_1 v_x)b+ak_2 v_x}dv_x \qquad (7)$$

where $\omega_0$ is Langmuir frequency $\omega_0^2=4\pi e^2/m_e$.

Expression (6) represents an insurmountable contradiction, since by definition $f_1(v_x)$ should not depend on $x,t$ (i.e. $a,b$), as well as the equation (7), indicating the absence of solutions $k_1, k_2\neq 0$ independent on $x,t$. However, this contradiction can be overcome by the introduction of energy-exchange ("collisional") terms on the right hand side of the kinetic equation (4). One can try for this to obtain some average values $k_1; k_2\neq 0$ over the wave period introducing the collision term $S$ on the right hand side of the kinetic equation with the dispersion equation of the form

$$1=\frac{-ab}{ak_1+bk_2}\cdot\omega_0^2\int\frac{\partial f_0(v)/\partial v_x}{(\omega-k_1 v_x)b+ak_2 v_x-A}dv_x \qquad (8)$$

taking

$$A\equiv ak_2 v_x; \quad S\equiv Af_1(v_x)e^{-k_2 x}. \qquad (9)$$

In this case the dispersion equation takes the form [1], [2]:

$$1=\frac{-a}{ak_1+bk_2}\int\frac{\partial f_0(v)/\partial v_x}{(\omega-k_1 v_x)}dv_x; \qquad (10)$$

$$\left(\frac{a}{ak_1+bk_2}\right)_{av}=\frac{1}{2\pi}\int_0^{2\pi}\frac{dy}{k_1+k_2\cdot\tan y} \qquad (11)$$

with the need of correction of Eq. (18) in [2] and

$$f_1(v_x)=\frac{e}{m_e}\frac{\partial f_0(v)/\partial v_x}{(\omega-k_1 v_x)}E_0, \qquad (12)$$

where all integrals can be taken including improper integrals (10) in the principal value sense or using cutting off $f_1(v_x)$ which leads to weak modulated Van Kampen beam (see also some constrains on $k_2$ in *Appendix 2*, [1]).

The required for finding averaged $k_1$ and $k_2$ the second integral energy conservation equation can be written, for example, as relation of the general type with arbitrary energy-exchange terms:

$$\int f_1(v_x) k_2 v_x \Delta\varepsilon(v_x) dv_x = \int f_1(v_x) \left[ n_e \sigma_1(v_x) v_x \Delta\varepsilon_2(v_x) + n_m \sum_{n>1} \sigma_n(v_x) v_x \Delta\varepsilon_{n+1}(v_x) \right] dv_x \qquad (13)$$

(with further integration over $v_x$) and with possible resonances of $k_2$ in dependence on $\omega$ and arising excited molecules where $\Delta\varepsilon_1(v_x)$ is energy loss per electron associated with a decrease in the number of electrons with an additional wave propagation velocity $v_x$; $\Delta\varepsilon_2(v_x)$, $\Delta\varepsilon_3(v_x)$ ..., are portions of transmitted energy; $\sigma_n(v_x)$ are collision cross sections; $n_e$ is electron density; $n_m$ is density of atoms and molecules. In this expression

$$\Delta\varepsilon_1 \sim \frac{m_e}{2}\left[(v_{xM} + v_x)^2 - v_{xM}^2\right] \qquad (14)$$

where $v_{xM}$ is proper velocity of electron along the axis $x$ in Maxwell distribution $f_0(v)$ independent on the wave speed. The expression (13) contains unknown both $k_1$ and $k_2$, and one can add into it also any other actual arbitrary energy exchange process terms including "collisional" and "collisionless" energy-exchange terms.

Note that the complex dispersion equation with the substitution of complex wave functions at enough large $k_2 \neq 0$ has nothing to do with the real dispersion equation of the real Eq. (7) and leads to collisionless damping in violation of the law of energy conservation [1] – [4].

## 3. Conclusion

The laws of nature are characterized as relations between the physically observable real quantities. Therefore, in some cases, the use of physical real variables can be regarded as testing and probing of calculations of physical processes made by using surprisingly fruitful huge mathematical apparatus of the theory of functions of complex variables.

In the case of real exponentially damping plasma waves in collisionless plasma with a given frequency, dispersion equation contains two parameters to be determined: the real wave number and the damping decrement, to find these you need to add the second equation, which can be the equation of energy conservation.

In the case of complex exponentially damping wave functions with complex wave number $k = k_1 - ik_2$ which causes nonlinear complexity of wave functions, complex dispersion equation is equivalent to the two equations correspondingly for the real and imaginary parts, determining the two parameters: the real wave number $k_1$, decrement/increment $k_2$, and two non-physically, unnaturally simultaneously coexisting exponentially divergent and damping solutions $k_1 \pm ik_2$ (the root is $(k_1, k_2^2)$) and besides more obviously violating the law of energy conservation, regardless of any other equations.

In the case of zero-decrement (i.e. the linear complexity approximation) complex wave functions lead to the real dispersion equation to be determined with the real wave number $k_1$ at $k_2 = 0$ (non-damping waves).

Analysis of the real dispersion equation for physically observable real values of the wave number and the decrement for the real wave functions being the real part of the complex wave functions shows that damping wave solution $k_1$, $k_2$ can exist only in the presence of terms of the collisional kinetic equation of a certain type, which corresponds to the energy conservation law with the inclusion of energy-exchange terms of very general form. More precisely, the exact solution still does not exist here, but it makes sense of the solution averaged over the period of oscillations.

It is carried out extended consideration of the problem of exponentially damping/growing real and complex plasma waves starting from a number of logical contradictions of collisionless complex nonlinear plasma damping/growing waves and divergence of electron distribution function at the singular points.

There is shown the logical fallacy of simple nonlinear extensions of linear complex expressions for the physical processes in the form of nonlinear complexity replacing the real values (for example, a real wave number or frequency in complex wave) by a complex wave number or complex frequency in the widespread paradigm of damping electron waves in collisionless plasma.

All this leads to the need to revise the widely represented for more than half a century works, including textbooks, the problem of collisionless damping of plasma waves and the use of basic plasma physics functional parameters such as the complex conductivity and the complex permittivity in the case of nonlinear complex extensions in complex expressions for the wave functions. Only real dispersion equation can be considered correct because it does not lead to the collisionless damping. Collisionless damping in the sense of Landau damping does not exist as a physical phenomenon. At the same time there remains an open problem of the possible use of expansions in the complex or real wave functions for obtaining the real dispersion equation for collisional plasma.

Conclusion about the impossibility of collisionless (ie, without entering into the kinetic equation of additional energy-exchange terms) damping/growing of plasma waves is general in nature, due to the law of energy conservation.

---

* **Footnote.** This topic below concerns the foundations of the large number of works in the last six decades on the collisionless damping of plasma waves.

Collisionless damping/growing is the trivial result of the using complex values $\omega$ and/or $k$ in arbitrary extending of the usual complex wave functions with real $\omega, k$. Fatal complex terms arise in the differentiation of complex nonlinear exponential wave functions with complex $\omega$ and/or $k$ and then appear in the denominator of the sought-for expression for the complex perturbation distribution function $f_1(v_x)$ as the sum of the real and imaginary terms, that leads to a complex dispersion relation with its virtual unphysical complex root $(k_1, k_2^2)$. Real dispersion equation for real wave functions is equivalent to the dispersion equation which is obtained by linear transformations of the linear, at real $k$ and $\omega$, complex wave functions (as expected in the case of one unknown $k$ at a given $\omega$). Nonlinearly complex wave solutions with complex $k$ or $\omega$ in collisionless plasma as a kind of independent mathematical objects should be damping/growing by definition, with no connection to any conservation laws. But along this, the real wave solutions of the real dispersion equation in such plasma should be always only non-damping/non-growing!

By the linear complexity it is implied that the use of the real part of the complex wave functions $\sim \exp[i(\omega t - kx)]$ for the perturbation of the distribution function $f_1(x,t)$ and respectively the real part of the electric field $E(x,t) \sim i\exp[i(\omega t - kx)]$ with the real $\omega, k$ leads to the same real dispersion equation when choosing, respectively, the imaginary parts of these wave functions. Thus, in this case there is no "intermixing" of the real and imaginary parts of the complex wave functions, they transform independently.

In the justification of collisionless damping, it is commonly used analytic continuation of the function $f_1(v_x)$ into the complex plane of velocity $v_x$ in the integral component of the dispersion equation

$$\int_{-\infty}^{\infty} f_1(v_x) dv_x \sim \int_{-\infty}^{\infty} \frac{\partial f_0(v)/\partial v_x}{\omega - kv_x} dv_x \ .$$

But physically natural cutting off function $f_1(v_x)$ (positivity condition of the complete distribution function) eliminates as a singularity and the need to checking analyticity of the integrand at its analytical continuation to the complex plane $v_x$ with real $k$. Complex dispersion equation can be obtained by going to the complex $k$ with the same problems of binomial complex root ($k_1, k_2$). Then the integral is approximately calculated as the sum of the real principal value and imaginary residuum (half-residuum?) at $v_x = \omega/k$ with complex $k$ (or, equivalently, complex $\omega$) and the need to prove analyticity (cf. also the more clear equivalent analysis in *Appendix 2*, [1]).

*The only right alternative to the finding both $k_1$ and $k_2$ (at a given real frequency $\omega$) from the solution of complex dispersion equation with the virtual non-physical binomial complex root $k_1 \pm ik_2$ is the way of solving the real dispersion equation with $k_1$ and $k_2$ combined with the real equation of energy conservation with $k_1, k_2$ leading to a completely different result for solution $k_1$ and $k_2$.*

"Landau damping" with its number of logical errors is the obvious unforgivable logical mistake! In addition to the general in principle inadmissibility of using complex dispersion equation, even usually given in the literature standard procedure for calculating the root $\pm k_2$ contains logical errors and leads to the 2 times underestimation of $k_2$ (if using half-residuum instead of the full residuum) value (see [1], pp.13 - 14).

Landau damping approximate solution corresponds to $|k_2| \ll k_1$ (with the relevant solving procedures inapplicable at possibly large unknown $|k_2|$), however obviously one can regard also a more general solution for the case $|k_2| \lesssim k_1$. But the use of such method of calculation $k_1, k_2$ in the case of large $|k_2|$ requires verification analyticity of the function $f_1(v_x)$ of complex $v_x$. Note that initial a priori assumption $|k_2| \ll k_1$ may be erroneous, even if it leads to the same end result. Note, that at solving complex dispersion equation decrement $k_2$ depends only on $f_0(v)$ and $\omega$, and does not depend on energy loss with violation of the energy conservation law. At the same time finding virtual non-physical solutions $k_1, k_2$ of the complex dispersion equation can represent only abstract mathematical interest.

The fallacy of "collisionless damping" of plasma waves is an elementary and at once mathematical consequence of the non-commutativity $\hat{D}\text{Re} F(\omega, k, x, t) \neq \text{Re}\hat{D} F(\omega, k, x, t)$ where $\hat{D}$ is operator of the procedure for obtaining the dispersion equation, and the wave function $F(\omega, k, x, t)$ has nonlinear complexity at a complex $\omega$ or $k$ and, moreover, the right and left parts can, contrary to the very sense of the dispersion equation, depend on $x, t$ correspondingly to some presumed view of the function $F$ and, moreover, differently for the right and left sides of the inequality (see *Appendix 2*).

*The very striking is ubiquity of traditional unthinking use of complex dispersion equations that are physically untenable in this case because of the violation the energy conservation law.*

# Landau damping: sampling of some classic textbooks and scholar articles[†]

[[†]Some arbitrary choice of these diverse works of different years has purely illustrative character and does not include numerous works both educational and with original calculations of the collisionless damping of plasma waves. It demonstrates not reduced and widespread interest to this problem.]

The continued interest in the decades to the paradoxical for the good sense mystical collisionless damping is caused by lack of the explicit mathematical formulation of the mechanism of energy flow away, although natural and intuitive explanation of its absence is just the possibility to ignore the energy-exchange ("collision terms" including radiation, collective, long-range and other types interactions with non-Landau damping) in the kinetic equation when opportunities arise to completely solve the dispersion equation in the case of complex, but not with the natural physical real $\omega, k$ wave functions which are some specific class of complex mathematical objects with their special properties.

This simple explanation eliminates the need for a conglomeration of sophisticated mathematical transformations including complex functions ("Landau damping"), which in fact do not lead to credible results, but lead to even more sophisticated new mathematical complex value refinements.

Uselessness and futility of these constructions is evident in the simple and unavoidable physically justified demand to give up the use of complex dispersion equations with their binomial roots, and use only the real dispersion two-parametric equations in which case there arises a need to add the second equation of energy conservation for finding these parameters. One can still in severe cases use expansions of the dispersion equation in the real individual wave functions with their individual real dispersion and energy balance equations.

## SUMMARY

(in principle impossibility of damping/growing wave solutions in plasma with collisionless kinetic equation and Maxwellian background)

In the case of collisionless plasma which is characterized by collisionless kinetic equation with the lack of collisional energy-exchange term in the right-hand side, complex linear wave function of the form $\exp[i(\omega t - kx)]$ with real $\omega$ and $k$ (or the linear sum of such complex wave functions) leads to the real dispersion equation (DE) with non-damping Vlasov waves. Extrapolation of DE to the complex $k = k_1 - ik_2$ leads to exponentially damping/growing waves with complex DE. It is elementary logical error since complex nonlinear DE with $k_2$ in both real and imaginary parts of DE has two complex roots $k = k_1 \pm ik_2$ simultaneously determining $k_1$ and $k_2$ without any connection with the energy conservation equation. At the same time, the use of equivalent wave function $\exp(\pm k_2 x)\cos(\omega t - k_1 x)$ leads to real DE with the real $k_1$ and $k_2$. To determine them, you need to add the second equation of energy conservation with $k_1$ and $k_2$, what is equivalent to adding the right-hand side of the kinetic equation with energy-exchange collisional terms.

This obviously means impossibility of collisionless damping/growing wave solutions in this collisionless sense plasma.

In this regard, it requires a fundamental revision of numerous plasma works using nonlinearly complex parameters such as complex plasma frequency or complex wave numbers. In this case, for physically based real expressions relating the plasma parameters, it is not enough to nullify imaginary parts of these parameters in the resulting complex expressions. It is necessary to repeat the derivation of these expressions using the original real values of these physical parameters in the equivalent real initial wave functions (cf. *Appendix 2*). Of course, these considerations do not apply to the cases when the real plasma wave functions are presented in the form of identical integral representations of Fourier or Laplace, or any other expansions with complex components but with finally real dispersion equation.

(In particular, as an example, this applies to the our work *Soshnikov V.N., Nedzvedsky V.Ya.* Damping of plasma sound in a weakly non-equilibrium plasma. "Plasma Physics" (1988), **14**, 16, p. 1248 (Russian), which calculated the second and the third iteration of the damping rate (decrement) of ion sound in the approximation of a collisionless Landau damping.)

The collection features of the unreliability of collisionless damping, refer to the school lectures:

**Werner Herr, CERN (2013), Lectures.**
**Physics of Landau Damping - CERN Accelerator School**
**http://cas.web.cern.ch/cas/Norway-2013/Lectures/Herr1.pdf**
**Landau damping - the mystery. First publication in 1946. Applied to longitudinal oscillations of an electron plasma. Was not believed for ≈ 20 years.**

______________________________________________________________